\documentclass[
    reprint,
    superscriptaddress,
    amsmath,
    amssymb,
    aps,
    prx,
    longbibliography
]{revtex4-1}

\usepackage{graphicx}
\usepackage{dcolumn}
\usepackage{bm}
\usepackage[colorlinks=true,urlcolor=blue,linkcolor=blue,citecolor=blue]{hyperref}
\usepackage[normalem]{ulem}
\usepackage{tikz}
\usetikzlibrary{quantikz}
\usepackage{amsmath}
\usepackage{amssymb}
\usepackage{bbold}
\DeclareMathOperator{\Tr}{Tr}

\newcommand{\customref}[2]{\hyperref[#1]{\ref*{#1}#2}}

\usepackage{xcolor}

\definecolor{Ured}{HTML}{cc0000}
\definecolor{Ublue}{HTML}{1f65cf}
\definecolor{Ugreen}{HTML}{198a11}

\begin{document}

\title{Avalanches and many-body resonances in many-body localized systems}

\author{Alan Morningstar}
\affiliation{Department of Physics, Princeton University, Princeton, NJ 08544, USA}

\author{Luis Colmenarez}
\affiliation{Max Planck Institute for the Physics of Complex Systems, Noethnitzer Stra{\ss}e 38, Dresden, Germany}

\author{Vedika Khemani}
\affiliation{Department of Physics, Stanford University, Stanford, CA 94305, USA}

\author{David J. Luitz}
\affiliation{Physikalisches Institut, University of Bonn, Nussallee 12, 53115 Bonn, Germany}
\affiliation{Max Planck Institute for the Physics of Complex Systems, Noethnitzer Stra{\ss}e 38, Dresden, Germany}

\author{David A. Huse}
\affiliation{Department of Physics, Princeton University, Princeton, NJ 08544, USA}
\affiliation{Institute for Advanced Study, Princeton, NJ 08540, USA}

\date{\today}

\begin{abstract}
We numerically study both the avalanche instability and many-body resonances in strongly-disordered spin chains exhibiting many-body localization (MBL). Finite-size systems behave MBL within the MBL regimes, which we divide into the asymptotic MBL {\it phase}, and the {\it finite-size MBL regime}; the latter regime is, however, thermal in the limit of large systems and long times.  In both Floquet and Hamiltonian models, we identify some ``landmarks" within the MBL regimes.  Our first landmark is an estimate of where the MBL {\it phase} becomes unstable to avalanches, obtained by measuring the slowest relaxation rate of a finite chain coupled to an infinite bath at one end. Our estimates indicate that the actual MBL-to-thermal phase transition occurs much deeper in the MBL regimes than has been suggested by most previous studies.  Our other landmarks involve system-wide many-body resonances:  We find that the effective matrix elements producing eigenstates with system-wide many-body resonances are enormously broadly distributed. This broad distribution means that the onset of such resonances in typical samples occurs quite deep in the MBL regimes, and the first such resonances typically involve rare pairs of eigenstates that are farther apart in energy than the minimum gap. Thus we find that the resonance properties define two landmarks that divide the MBL regimes of finite-size systems into three subregimes: (i) at strongest randomness, typical samples do not have any eigenstates that are involved in system-wide many-body resonances; (ii) there is a substantial intermediate subregime where typical samples do have such resonances, but the pair of eigenstates with the minimum spectral gap does not, so that the size of the minimum gap agrees with expectations from Poisson statistics; and (iii) in the weaker randomness subregime, the minimum gap is larger than predicted by Poisson level statistics because it is involved in a many-body resonance and thus subject to level repulsion. Nevertheless, even in this third subregime, all but a vanishing fraction of eigenstates remain non-resonant and the system thus still appears MBL in most respects. Based on our estimates of the location of the avalanche instability, it might be that the MBL {\it phase} is only part of subregime (i), and the other subregimes are entirely in the thermal phase, even though they look localized in most respects, so are in the finite-size MBL regime.
\end{abstract}

\maketitle

\section{Introduction\label{sec:introduction}}

\begin{figure}
	\includegraphics[width=1.0\linewidth]{{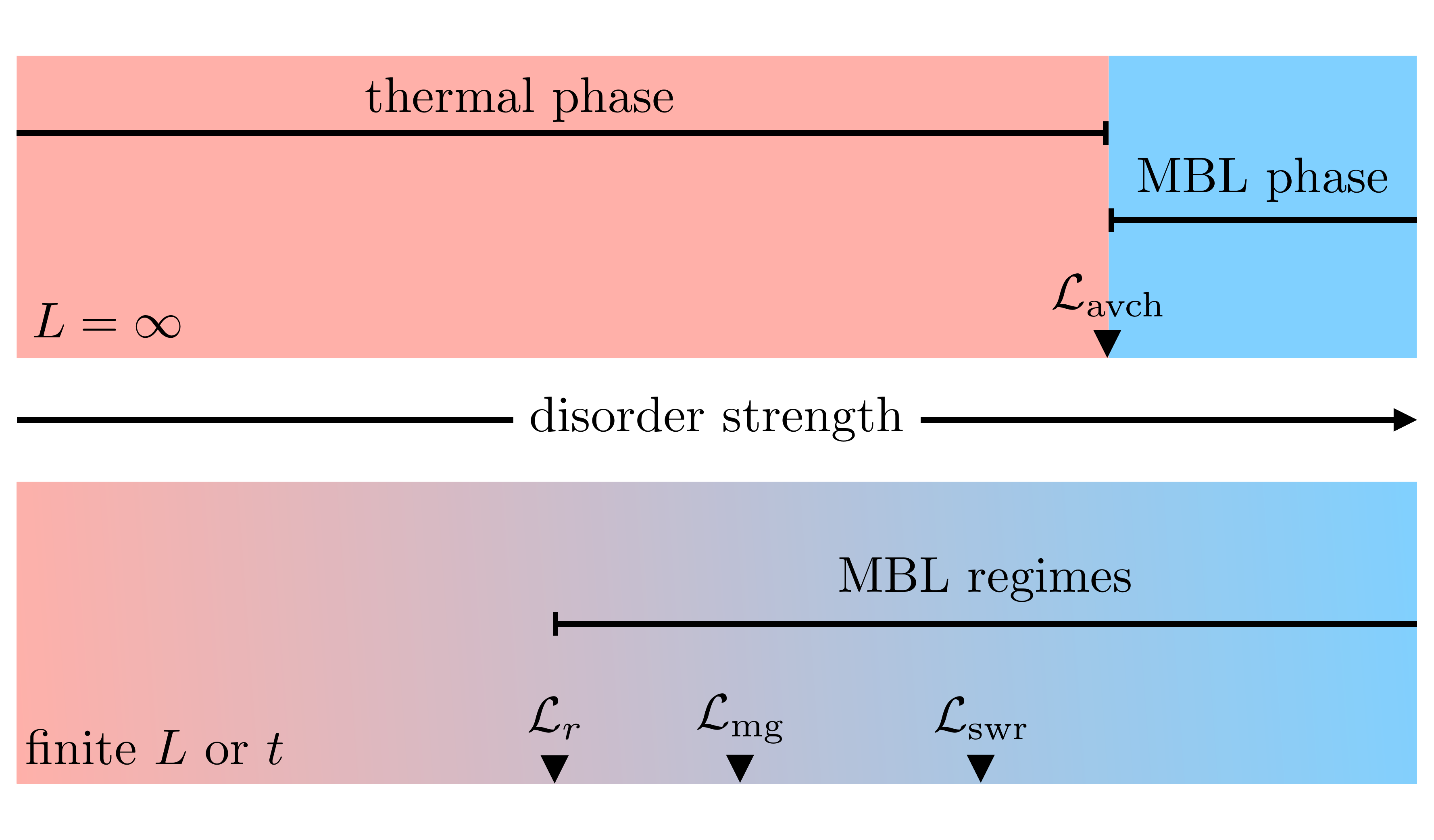}}
	\caption{Sketch of the MBL phase diagram. \textbf{Top:} In the infinite size limit, the transition out of the MBL phase is believed to occur due to an instability towards the formation of thermalizing avalanches. The disorder strength at which this occurs marks the $L=\infty$ limit of one of our ``landmarks", denoted $\mathcal{L}_\text{avch}$. \textbf{Bottom:}  At accessible finite system sizes $L$ and/or finite times $t$, we observe the significantly larger MBL {\it regimes}. The finite-$L$ crossover in the mean spectral gap ratio $\langle r \rangle$ from random matrix to Poisson statistics is one convenient landmark, denoted $\mathcal{L}_r$, marking the crossover from thermal behavior into the MBL regimes; $\langle r \rangle$ probes the level repulsion between all neighboring energy levels. The MBL regimes contain a few additional landmarks that probe the behavior of rare many-body resonances in eigenstates. At strong disorder exceeding a threshold marked by $\mathcal{L}_\text{swr}$, there are no system-wide resonances (swr) in any of the eigenstates of typical samples. For weaker disorder, a small number of eigenstates display system-wide resonances, but the {\it minimum gap} (mg) in the spectrum is non-resonant and shows negligible level repulsion. The level repulsion in the minimum gap appears at even weaker disorder within the MBL regime, at a landmark denoted by $\mathcal{L}_\text{mg}$. 
	Any of these finite-system landmarks that occur in the finite-size MBL regime will drift towards the true transition as $L$ and $t$ are taken to infinity, and some might drift past the transition and end up within the asymptotic MBL phase.
		\label{fig:phases_regimes_landmarks}}
\end{figure}

Many-body localized (MBL) systems fail to reach thermal equilibrium under their own dynamics, and have been a subject of intense interest over the last decade~\cite{Anderson1958,gornyi_interacting_2005,Basko-Altshuler2006,Oganesyan-Huse2007,Nandkishore-Huse2015,alet_many-body_2018,AbaninRMP2019}. They display a form of emergent integrability, characterized by the presence of an extensive set of local integrals of motion (``l-bits")~\cite{Serbyn-Abanin2013,Huse-Oganesyan2014}. The phenomenology of l-bits explains many distinctive features of the MBL phase, including its unusual slow dynamics~\cite{Znidaric-Prelovsek2008, Bardarson-Moore2012,Serbyn-Abanin2013b, Serbyn_2014, Khemani-Sondhi2015}, and the possibility for novel forms of ``localization protected" order in individual highly excited many-body eigenstates~\cite{Huse-Sondhi2013,Pekker_HilbertGlass, Chandran-Sondhi2014, Khemani_2016, Parameswaran-Vasseur2018}. The phase transition between an MBL and a thermal phase is a new class of dynamical phase transition, a complete understanding of which has thus far proved to be notoriously elusive: analytical treatments are mainly tractable only under phenomenological frameworks~\cite{Vosk-Altman2013,Vosk-Altman2015,Potter-Parameswaran2015, Dumitrescu-Vasseur2017,Thiery-DeRoeck2017b, Goremykina-Serbyn2018,Dumitrescu-Vasseur2019,Morningstar-Huse2019,Morningstar-Imbrie2020} and numerical simulations are restricted to very small system sizes that do not exhibit the asymptotic physics of large systems~\cite{Oganesyan-Huse2007,Znidaric-Prelovsek2008,Pal-Huse2010,Kjall-Pollmann2014,luitz_many-body_2015,Khemani-Huse2017, Khemani-Huse2017b}. Thus an interesting complementary approach to the transition has been to try to understand mechanisms by which MBL can be destabilized under certain conditions, and to then build numerical evidence for such mechanisms, and develop corresponding phenomenological models to capture the large-scale consequences of those mechanisms. This has been exhibited by a body of work that proposed  ``avalanches" seeded by rare Griffiths regions as a mechanism for destabilizing MBL in disordered systems ~\cite{DeRoeck-Huveneers2017,Thiery-DeRoeck2018}, attempted to numerically observe certain features of avalanches in minimal toy models ~\cite{Luitz-DeRoeck2017,goihl_exploration_2019, Crowley-Chandran2020}, and determined the consequences of an avalanche-driven transition in phenomenological renormalization group (RG) treatments~\cite{Thiery-DeRoeck2017b,Dumitrescu-Vasseur2019,Morningstar-Imbrie2020}. There have also recently been experiments investigating isolated avalanches in cold atomic systems~\cite{Leonard-Greiner2020}.

In this paper, we make a distinction between the MBL {\it phase}, which is the part of the phase diagram where the system remains MBL in the limits of an infinite system and infinite time, and the {\it finite-size MBL regime}, which is the part of the phase diagram for an accessible  finite-size system (and/or finite time scale) where this size system behaves MBL in most respects, although larger systems do thermalize.  When we say ``MBL regimes'' (plural) this means both the MBL phase and the finite-size MBL regime.

Our work advances our understanding of both (i) the phase transition out of the MBL phase, and (ii) the properties of many-body resonances in the MBL regimes. On (i), we present a qualitatively new way of estimating a bound on the boundary of the MBL phase defined by the avalanche instability.  This defines the first of three ``landmarks" that we estimate within the MBL regimes in this paper. Despite much work on avalanches, the theory has only started to be integrated with numerical simulations of microscopic systems in order to produce an estimate of where the avalanche instability occurs~\cite{Varma-Pekker2019}, so our work, along with subsequent work by Sels~\cite{sels_markovian_2021}, advances the state of the art in this direction. Our bound on the avalanche instability is at significantly larger disorder strengths than previous finite-size estimates of the MBL transition~\cite{devakul_early_2015,doggen_many-body_2018,sierant_polynomially_2020}. On (ii), we identify two additional system-size-dependent landmarks within the MBL regimes pertaining to the onset and nature of rare, long-range resonances in many-body eigenstates.
Our estimates for all three landmarks are well into the MBL regimes, but it may be the case that the latter two landmarks (pertaining to resonances) are, for all system sizes, not within the MBL phase.  See Fig.~\ref{fig:phases_regimes_landmarks} for a schematic representation of the various landmarks and the distinction between the MBL phase and regimes.

Our investigations are focused on regimes that were previously thought to be rather ``deep" in the MBL phase. We believe that, by doing so, we are beginning to try to remedy a methodological error that has been made by much of the MBL research community, the majority of the present authors included.  This error was to focus so strongly on the MBL phase transition before more thoroughly studying the MBL phase itself.  The MBL phase, and more generally Anderson localized phases, are gapless critical phases, with slow dynamics due to resonances and near-resonances, as emphasized early on by Mott~\cite{Mott1968,Gopalakrishnan-Huse2015, Serbyn_2014, Khemani-Sondhi2015}.  

The thorough study of the MBL phase appears to have been delayed partially because the description of that phase in terms of l-bits~\cite{Serbyn-Abanin2013,Huse-Oganesyan2014} superficially seemed rather simple and complete, and explained many features of the phase.  But that is incorrect, since those descriptions did not fully address the nontrivial dynamical properties of resonances and avalanches in the MBL phase.  And since the existence of the MBL phase is a dynamical phenomenon, a description that neglects important aspects of its dynamics is certainly incomplete. One conclusion of the present work is that the full l-bit description must include a lot of detailed structure that has generally been ignored in previous work.  Any many-l-bit process, and how it can be driven by any local operator, is a property of the l-bits.  Thus all the structure of all many-body processes that can flip any number of l-bits must be properly encoded in the details of the definitions of the l-bits and the l-bit Hamiltonian; previous work generally assumed the couplings in the l-bit Hamiltonian are essentially random, ignoring any such detailed structure and thus neglecting the strong many-body resonances.  In this paper we begin to explore some aspects of this many-body structure. We are not suggesting that the l-bit description of the MBL phase fails, only that such a description needs to capture many fine details of the system in order to be dynamically correct.

We also note that recently there have been a number of papers expressing various levels of skepticism about the stability of the MBL phase in the limits of large systems and very long times~\cite{weiner_slow_2019,Suntajs-Vidmar2020,Suntajs-Vidmar2020b,KieferEmmanouilidis-Sirker2021,SelsPolkovnikov2020,VidmarMierzejewski2021,Sels-Polkovnikov2021}, and a number of challenges to those conclusions~\cite{Abanin-Vasseur2021,Sierant-Zakrzewski2020,Panda-Znidaric2020,Luitz-BarLev2020,Ghosh-Znidaric2021}. While our results do not support (nor directly contradict) any of these arguments for or against the existence of the MBL phase, our work was, in part,  motivated by these works, which certainly have demonstrated that our understanding of the MBL regimes, transition, and crossovers is still rather incomplete.

Our finding that the MBL phase transition actually occurs very deep in the MBL regimes, and far from the numerically-accessible crossover between the finite-size MBL regime and thermalization, reinforces the idea~\cite{Crowley-Chandran2021} that the physics of this crossover is likely quite different from that of the ultimate phase transition.  This suggests that this crossover should probably be studied as a distinct phenomenon from the MBL phase transition.  For example, this crossover occurs quite generally in MBL systems in higher dimensions and with longer-range interactions, while the MBL phase transition is suppressed due to the avalanche instability in those other cases \cite{DeRoeck-Huveneers2017}. Another related point is that the arguments against many-body mobility edges~\cite{DeRoeck-Schiulaz2016} apply only to the MBL phase transition, and not to this finite-size MBL regime to thermal crossover, which very clearly shows apparent mobility edges in numerics~\cite{luitz_many-body_2015}.

The rest of the paper is organized as follows. In Sec.~\ref{sec:landmarks} we elaborate on the various landmarks mentioned earlier, that we study in depth later on. We summarize the avalanche argument and explain our strategy for locating where this instability destabilizes the MBL phase. We also explain how we think about many-body resonances in this work, and provide an overview of how we detect them. In Sec.~\ref{sec:models} we detail the concrete spin-1/2 models we use as the basis of our calculations: one is a new Floquet random-circuit MBL model in 1D with no extensive conserved quantities, while the other is the Hamiltonian ``standard model'' of MBL, the random-field Heisenberg chain.  We also include details of how we couple each system to a model infinite bath at one end, which we use to numerically bound the avalanche instability. In Sec.~\ref{sec:results_open} we begin our study by investigating our first landmark, where the avalanche instability destabilizes the MBL phase, using the dynamics of our spin models coupled to infinite baths at one end. Before estimating the other two landmarks, in Sec.~\ref{sec:untangle} we develop a method for ``undoing" the level repulsion between two eigenstates, thereby obtaining an estimate of a matrix element responsible for producing many-body resonances in eigenstates. Equipped with this tool, in Sec.~\ref{sec:results} we study the properties of isolated spin chains in order to understand the two remaining landmarks pertaining to rare, long-range, many-body resonances. Finally, we summarize and discuss our findings.

\section{Landmarks in the MBL regimes\label{sec:landmarks}}

In this section we elaborate on the ideas and necessary background related to the three landmarks we study in this work. 

\subsection{Avalanche instability}
\label{sec:avalanche_intro}

One of the more accepted theories of what drives the \textit{asymptotic} MBL phase transition in systems with quenched randomness and short-range interactions is the so-called ``avalanche instability"~\cite{DeRoeck-Huveneers2017}. It proposes that, at weak enough disorder, small locally thermal rare regions make MBL unstable by seeding an avalanche of ergodic regions that drives the system to thermalize. In the avalanche theory, the rate at which a naturally occurring thermal bubble thermalizes its localized surroundings is what determines how much the bubble grows. This can be understood concretely by thinking of a spin chain with random local fields. In an infinite sample there are (arbitrarily long) rare regions where, just by chance, the random fields are small and the system locally thermalizes. This results in a finite local bath that then tends to thermalize the nearby typical localized regions.  The  spins that are at a distance of $\ell$ spins away from this rare region are, if the avalanche does reach them, typically thermalized at a rate $\sim k^{-\ell}$, where $k$ is a number that increases as one goes deeper in to the MBL phase. Once the avalanches due to this thermal region have proceeded to distance $\ell$ in both directions, then it has thermalized a total of $N+2\ell$ spins, where $N$ is the (pre-avalanche) number of spins in the thermal rare region.  Thus the many-body level spacing of this now-enlarged thermalized region is $\sim 2^{-(N+2\ell)}$.  As long as this level spacing is smaller than the spin's thermalization rate, then the spin does see the thermal region as a reservoir with an effectively continuous spectrum and does get entangled with it.  The avalanche will stop ($\ell$ will stop growing) when these two energies become equal so the spin can ``see'' that the spectrum of the thermal region is really discrete, namely when $k^{-\ell}\sim 2^{-(N+2\ell)}$.  For $k<4$ this never happens, so the avalanche does not stop and the full system slowly thermalizes.  Thus, by this mechanism, the phase transition out of the MBL phase happens at $k=4$, yielding a critical thermalization rate that scales as $4^{-\ell}$. A similar scenario has been numerically verified in a minimal toy model comprising a thermal bubble (modeled by a random matrix) interacting with a chain of decoupled localized spins (model l-bits)~\cite{Luitz-DeRoeck2017}.

One of our goals in this work is to numerically study the avalanche instability in more realistic microscopic models of MBL, in which the emergent l-bits interact and are not known {\it a priori}. Interactions between l-bits means that a given spin some distance from the thermal bubble can couple to the bubble via many distinct processes, which makes it challenging to extract a single thermalization rate for the spin.  In this work, we present a new approach to this challenge of computing a thermalization rate in a realistic MBL system. There is no simple answer that we are aware of in the case of isolated systems, however when a finite system is coupled at one end to an infinite bath, and hence viewed as an open system with dynamics described by a (Lindblad or Floquet) superoperator, the thermalization rate can be seen as the inverse of the time scale on which the system reaches the (unique and always thermal) equilibrium steady state. The closest eigenmode to the steady state of the superoperator encodes this time scale. Hence, the superoperator eigenmodes provide a direct way to estimate thermalization time scales for MBL systems.  Note that this open system calculation models only one of the two avalanches that are spreading in both directions from a large locally thermalizing rare region within the bulk of a nominally infinite system.  The assumption is that these two avalanches do not directly affect each other, except through their effect on the thermal bubble's many-body density of states.  In Sec.~\ref{sec:results_open} we show that as the parameters of a finite MBL system coupled to an infinite bath at its end are varied, there is a point at which the thermalization time scale crosses through a $\sim 4^{-L}$ scaling ($k=4$); we interpret this as a finite-size estimate of the avalanche instability driven MBL transition in the corresponding isolated system. We assume that the avalanche instability is what asymptotically drives the transition out of the MBL phase of a disordered system. Hence, we expect this landmark measuring the onset of the avalanche instability to be a better estimate of the true MBL transition in the limit of large $L$. This is how we define the landmark denoted by $\mathcal{L}_\text{avch}$.

A closed-system approach to this problem was proposed in Ref.~\cite{Varma-Pekker2019}, which used a Wegner-Wilson flow method to extract l-bits and compute spatial decay rates of various correlation functions.  However, to our knowledge, none of the closed-system correlation functions examined in Ref.~\cite{Varma-Pekker2019} are directly related to the relaxation rate of a distant spin due to a bath or thermal bubble, which is the quantity that is needed to make an estimate of the avalanche instability.

\subsection{Appearance of system-wide resonances}
\label{sec:resonance_intro}

Further landmarks ($\mathcal{L}_\text{swr}$ and $\mathcal{L}_\text{mg}$) within the MBL regimes are related to rare many-body resonances.
We should therefore provide some clarification on how we use the term ``many-body resonance", which has been used variously to describe many physical scenarios ranging from resonances between thermal blocks in phenomenological RGs to isolated ``Mott-like" resonances in many-body eigenstates~\cite{Potter-Parameswaran2015, Vosk-Altman2015, Geraedts-Regnault2016, Bukov-Polkovnikov2016,Bulchandani-Gopalakrishnan2021}. In the present work, an eigenstate of the dynamics is ``many-body resonant" if it is a superposition of localized states that differ substantially in extensively many local regions, and the range of the resonance is the distance over which these local differences occur. A dynamical implication of this is that if the system is initialized in one of those localized states, it will tunnel to the other(s) under the dynamics. A ``many-body resonance" refers to a set of states that are related by this definition, for example, eigenstates that are all superpositions of the same set of localized states. The importance of these resonances has been discussed in various contexts, including the dynamical a.c. response of MBL systems~\cite{Gopalakrishnan-Huse2015} and theories of finite-size crossovers between MBL and thermalizing systems~\cite{Geraedts-Regnault2016, Khemani-Huse2017, Khemani-Huse2017b, Villalonga-Clark2020,Crowley-Chandran2021}.

In this work, we study system-wide many-body resonances rather deep in the MBL regimes, where these resonances are rare and each such resonance typically involves only two eigenstates.  We examine the properties of these resonant eigenstates from two perspectives:

From the first perspective, the structure of entanglement in the eigenstates of the dynamics is a direct probe of many-body resonances, as is level repulsion in the spectrum. We find that even deep in the MBL regimes, there is still residual level repulsion to be understood, including rare strongly repulsive pairs (resonances) in an otherwise Poisson-like spectrum. One way to pick out a rare system-wide resonance in an MBL system is to find the eigenstate with the most quantum mutual information between its end spins. A significant amount (compared to one bit) of quantum mutual information between end spins is an indicator that this eigenstate is participating in a two-state (or few-state) system-wide many-body resonance, and this has resulted in it being a Schr{\"o}dinger cat-like state \cite{edge_modes}. Indeed this is one of the measures we use in order to identify rare, isolated system-wide many-body resonances in eigenstates in the MBL regimes. This identifies a system-size dependent landmark $\mathcal{L}_\text{swr}$ that separates the MBL regimes into: a stronger-randomness regime where typical samples have no such resonances and the probability that a sample has such a resonance is decreasing with increasing $L$, and a weaker-randomness regime where the number of such resonances per sample increases (exponentially) with increasing $L$.

We find another landmark $\mathcal{L}_\text{mg}$ by examining the amount of level repulsion present in the minimum gap (mg), {\it i.e.}, between the two eigenvalues of the dynamics that are nearest to each other. Poisson statistics predicts that the smallest gap in the Floquet spectrum of a sample is on average $2 \pi / 4^L$ ($2^L$ times smaller than the average gap between all adjacent eigenvalues), and hence comparing the smallest gap to this prediction reveals a landmark at which the minimum gap begins to typically undergo significant level repulsion.

We note that the minimum gap bears additional theoretical relevance, as assumptions on its scaling are a building block in the proof for the stability of MBL~\cite{imbrie_many-body_2016,imbrie_diagonalization_2016}, and our analysis confirms that the assumption of ``limited level attraction" is indeed comfortably valid for the system sizes we can test.

From the second perspective, the entanglement in, and level repulsion between, eigenstates is a result of off-diagonal matrix elements of the Floquet operator or Hamiltonian that couples states that are more localized than the eigenstates, and that differ extensively. Thus appropriate off-diagonal matrix elements can also be studied in order to learn about many-body resonances in MBL systems. If we are able to ``undo" some of the entanglement and level repulsion by rotating away from the basis of eigenstates back towards a less entangled basis (closer to the computational basis), then off-diagonal matrix elements of the Floquet operator or Hamiltonian in that new basis can be considered to be the source of the level repulsion---as these off-diagonal matrix elements get rotated away, the energies get pushed apart and the states get more entangled. Thus these matrix elements characterize the underlying resonance. In Sec.~\ref{sec:untangle} we develop a useful tool in this spirit that allows us to associate an off-diagonal matrix element that characterizes the strength of the level repulsion between any two eigenstates. 
When the two eigenstates are both cat-like superpositions of two more localized states, we are able to retrieve a matrix element that is larger than or comparable to the corresponding gap.
When the two eigenstates are not resonant, then the matrix element is very small in comparison to the gap. This resonance criterion is similar to the criterion introduced in Ref. \cite{Serbyn-Abanin2015}, but here we use a different approach to determine the relevant matrix element. Using this procedure for characterizing resonances with these matrix elements, we are able to estimate $\mathcal{L}_\text{swr}$ and $\mathcal{L}_\text{mg}$ in a second, independent way.

After characterizing the MBL regimes with these matrix elements, the picture that emerges is as follows:  The distribution of matrix elements is broad on a log scale, and the {\it typical} (or median) matrix element between eigenstates scales as $\sim k_t^{-L}$, where $k_t>2$ is a number that increases (without limit) as we go deeper (to stronger randomness) into the MBL regimes. Note that we do find that the apparent $k_t$ is $L$-dependent for the sample sizes that we can study numerically.  Extremely deep in the MBL regimes, $k_t$ is large enough that there are typically no system-wide resonances in the many-body spectrum. At weaker disorder, marked by the threshold $\mathcal{L}_\text{swr}$, a vanishing fraction of eigenstates begin to be involved in system-wide resonances. These isolated resonances involve atypically large matrix elements in the tails of the distributions. However, as long as $k_t>4$,  the typical matrix element is small compared to the expected minimum gap in the spectrum ($\sim 4^{-L}$) and level repulsion of that minimum gap is still typically negligible.  In other words, in this intermedate regime there are eigenstates with rare system-wide many-body resonances because of atypically large matrix elements, even though the minimum gap is also typically not involved in a resonance. At even weaker-randomness within the finite-size MBL regime where $k_t<4$, the minimum gap is involved in a resonance and decreases with $L$ more slowly than predicted by Poisson level statistics. This distinction allows us to define the landmark $\mathcal{L}_\text{mg}$, where $k_t=4$ and the minimum gap changes from Poissonian to non-Poissonian.

\subsection{Summary of landmarks}
For later reference, here is a list of our ``landmarks'', the symbols with which they are represented (see also Fig.~\ref{fig:phases_regimes_landmarks}), and how they are estimated with finite-$L$ data:
\begin{itemize}
    \item $\mathcal{L}_\text{avch}$: This is where the avalanche instability destabilizes the MBL phase. We estimate it by where the thermalization rate in our open systems is scaling with $L$ as $\sim 4^{-L}$.
    \item $\mathcal{L}_\text{swr}$: This is where the number of system-wide many-body resonances per sample changes from asymptotically zero at large $L$ to a number that instead grows with increasing $L$.  We estimate it by where the number of such resonances per sample is not changing with $L$.
    \item $\mathcal{L}_\text{mg}$: This is where the minimum spectral gap in a sample begins/ceases to behave as expected in a Poisson spectrum with no level repulsion.  We estimate it by where the mean minimum gap is scaling with $L$ as $\sim 4^{-L}$.
    \item $\mathcal{L}_r$: This is the ``conventional'' boundary of the MBL regimes marked by the finite-size crossing of the mean level spacing ratio $\langle r \rangle$.
\end{itemize}
In the above list we have also included the reference point $\mathcal{L}_r$ for convenience, since we will need to refer to this point too; the three new landmarks within the MBL regimes that we study are the first three on this list.

Note that since we are only able to access small system sizes, we will treat these estimated landmarks as $L$-dependent quantities. Of course their asymptotic locations as $L\to\infty$ are of great interest, but we are not in the asymptotic regime, so we assert that estimates at accessible $L$ are still meaningful and help us to better understand the finite-size MBL regimes and potentially the MBL phase.

One notable feature of our results is that all of these landmarks exhibit similar strong finite-size effects for the size ranges we can study.  This suggests that the strong finite-size effects in the level statistics that have been widely studied may not be due to physics that is special to that thermal-to-finite-size-MBL crossover, but the same physics may also be producing strong finite-size effects much deeper in the MBL regimes.  At this point there does not seem to be any concrete and plausible theoretical understanding of these finite-size effects, so we have no guidance in how to extrapolate our estimates of these landmarks to the limit of infinite systems.  However, since they all move monotonically to stronger randomness with increasing system size, our numerical estimates should be reliable lower bounds on the randomness that these landmarks go to in the large-system limit.

Now that we have laid out the main ideas of this work, in the next section we present the models we use in later sections to elaborate on these ideas.

\section{Models\label{sec:models}}

We consider two models of MBL in this work: one is a time-periodic (Floquet) quantum circuit that we introduce, similar to circuits considered in Refs.~\cite{sunderhauf_localization_2018,garratt_many-body_2020}, and the other is the standard random-field Heisenberg (XXX) Hamiltonian model~\cite{Znidaric-Prelovsek2008, Pal-Huse2010, luitz_many-body_2015}. We include the Hamiltonian model to ensure all of our conclusions are consistent across the Floquet and Hamiltonian cases, and to make contact with previous work, but we believe our Floquet model is advantageous in several respects that we detail in Sec.~\ref{sec:Floquet}. Both models govern the unitary dynamics of a one-dimensional system of $L$ qubits (sites). While in the Hamiltonian model both the total energy and total $Z$ magnetization are conserved, the Floquet model has no conservation laws. Both of these models are designed such that in the limit of strong disorder the eigenstates of the dynamics are Fock states of the Pauli $Z$ operators on the $L$ sites. We also refer to this basis as the computational basis.

In addition, we extend the unitary models by coupling each to an infinite quantum bath at the left end of the system in order to study the avalanche instability. This is achieved in the case of the Hamiltonian model by introducing a complete operator basis of three nontrivial Lindblad jump operators on the first site. This bath relaxes both of the conserved quantities, and corresponds to infinite temperature and zero field. For our Floquet circuit we use a Floquet superoperator with a generic dissipative action on the first site.  
MBL systems are unstable when coupled to an infinite thermal bath~\cite{levi_robustness_2016,wybo_entanglement_2020,goihl_exploration_2019,lenarcic_critical_2020}.  
But the dependence of the rate of thermalization of the farthest spin on the system length $L$ in this setting indicates whether or not an infinite MBL system is stable to avalanches initiated by a large but finite bath.
So here we are particularly interested in the rate at which the system thermalizes due to the infinite bath, and focus on the decay rates of the slowest decay mode, given by the eigenvalue of the superoperators closest to the steady state eigenvalue $\sigma=1$ in the Floquet case, and $\lambda=0$ in the Lindblad case. Throughout this paper we use open boundary conditions in order to access the largest range of distances within the systems.

\subsection{Floquet random circuit \label{sec:Floquet}}

We introduce a one-dimensional, time periodic, random unitary circuit which can exhibit MBL. The model consists of two types of random unitary gates: one-site gates $d_i$, and two-site gates $u_i$ coupling site $i$ and $i+1$. Tuning the strength of the two-site gates drives the model through a MBL transition~\cite{sunderhauf_localization_2018}.

The one site gates $d_i$ are generated by sampling, for each site $i$, a $2\times 2$ random matrix from the circular unitary ensemble (CUE) and then diagonalizing it. This means that for each realization of the circuit we choose the computational $Z$ basis to be the eigenstates of all of the $d_i$. The distribution from which we sample the two-site gates is invariant under this choice, so this is a matter of convenience.

The two site gates $u_i$ act on site $i$ and $i+1$ and are obtained from 
\begin{equation}
    u_i = \exp\left(\frac{\mathrm{i}}{\alpha} M_i\right) \in \mathbb{C}^{4\times4},
    \label{eq:twositegate}
\end{equation}
where $1/\alpha$ controls the interaction strength ($\alpha$ is the relative disorder strength), and $M_i\in \mathbb{C}^{4\times 4}$ is a random matrix sampled from the Gaussian unitary ensemble (GUE). From these building blocks, we create the Floquet unitary by first applying a layer of all of the one-site gates given by $U_d = d_1 \otimes d_2 \otimes \dots \otimes d_L$, and then applying the two-site gates in an order given by a random permutation $\pi\in S_{L-1}$:
\begin{equation}
    U_u = \prod_{i=1}^{L-1} \text{mat}(u_{\pi(i)}),
    \label{eq:Uu}
\end{equation}
where $\text{mat}(u_{i}) = \mathbb{1}_{2^{i-1}} \otimes u_i \otimes \mathbb{1}_{2^{L-i-1}}$ is the matrix representation of the gate $u_i$ in the full Hilbert space. There is no gate connecting site $1$ and site $L$, since we use open boundary conditions.
The resulting random circuit is then exemplified by

\begin{equation}
    U = U_u U_d = \begin{tikzpicture}[baseline=(current bounding box.center), scale=1]
        \foreach \x in {0,0.5,1,1.5,2,2.5,3}
        {
            \draw[thick] (\x, -0.6) -- (\x, 3);
        }

        \foreach \x/\y/\name in {0/2.5/1,0.5/0.5/2,1/1/3,1.5/1.5/4,2/0/5,2.5/2/6}
        {
            \draw[thick, fill=Ured,rounded corners=2pt] (\x-0.1,\y) rectangle +(0.7,0.4); 
            \draw (\x+0.25,\y+0.2) node {$u_{\name}$}; 
        }

        \foreach \x/\name in {0/1,0.5/2,1/3,1.5/4,2/5,2.5/6,3/7}
        {
            \draw[thick, fill=Ugreen,rounded corners=2pt] (\x-0.2,-0.5) rectangle +(0.4,0.4); 
            \draw (\x,-0.3) node {$d_{\name}$}; 
        }
    \end{tikzpicture}.
    \label{eq:circuit}
\end{equation}
The random permutation of gates removes the intrinsic difference in even and odd system sizes present in brickwork circuits and is convenient to treat even and odd system sizes on the same footing.

We use this Floquet model because we think it has enough advantages over the standard Hamiltonian model of Section~\ref{sec:Hamiltonian} to justify its introduction. 
This circuit model is designed to be free of any conservation laws. This means that we do not have to consider how the physics of conserved quantities like energy or particle number interacts with the physics of MBL. 
It also means we can treat every eigenstate of the dynamics on equal footing and study statistics over eigenstates as a function of solely the disorder strength, without also having to resolve their dependence on any conserved quantities. In our quantum circuit we use gates drawn from isotropic distributions of random matrices, so we also avoid choosing a special basis, which is an attractive property when studying universal aspects of quantum dynamics~\cite{Nahum-Haah2017,Khemani-Vishwanath2018}. As mentioned above, for each realization we do align our computational basis for each site with the eigenstates of the corresponding single-site gate, but the two-site gates are still drawn from a distribution without a special direction even after this alignment. 

We also note that the locality of this quantum circuit implies that the computational cost of applying it to a state is $O(L 2^L)$. Thus while the Floquet unitary does not have a sparse matrix representation, it can be applied one gate at a time, and so it is compatible with algorithms that rely on matrix-free matrix-vector products like geometric sum filtering~\cite{Luitz2021}, which we use to access large system sizes.

\subsection{Hamiltonian \label{sec:Hamiltonian}}

We also carry out our study on the standard random-field Heisenberg model. The Hamiltonian of this model is
\begin{eqnarray}
\label{eq:xxz_hamiltonian}
    H = \frac {1}{4} \sum_{i=1}^{L-1} \vec{\sigma}_i \cdot \vec{\sigma}_{i+1} + \frac{1}{2} \sum_{i=1}^L h_i Z_i,  
\end{eqnarray}
where $\vec{\sigma}_i = (X_i, Y_i, Z_i)$, $h_i$ are independent samples of a uniform random variable on $[-W, W]$, and $W$ is the parameter that tunes the disorder strength. The total magnetization $M=\sum_i Z_i$ is conserved so we restrict ourselves to the sector $M=0$ for even $L$ when we study isolated systems.

\begin{figure}
\includegraphics[width=1.0\linewidth]{{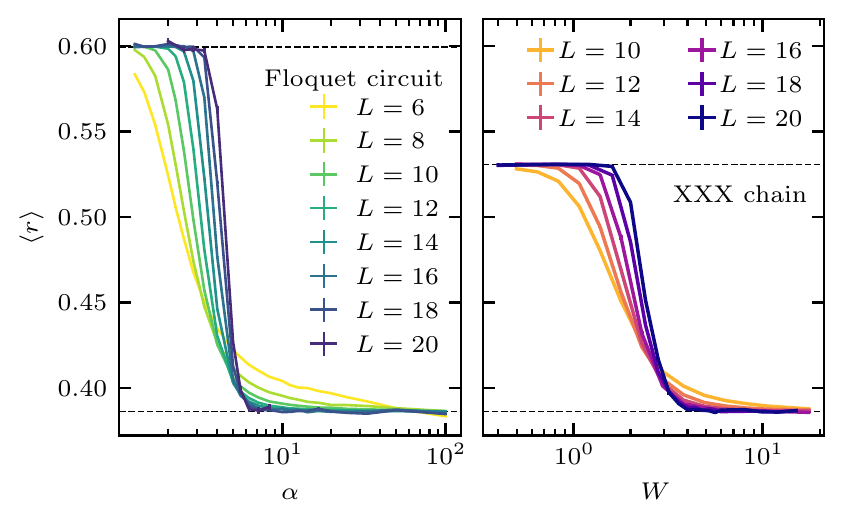}}
\caption{Mean level spacing ratio $\langle r \rangle$ as a function of disorder. The finite-size crossings drift to larger disorder as $L$ is increased.  For the largest $L$ the crossings are at $\alpha \cong 5.9$ for the Floquet model (\textbf{left}) and at $W \cong 3.1$ for the  Hamiltonian model (\textbf{right}).
Error bars are 68\% bootstrap confidence intervals here and in all other figures, but are too small to see here.  The Floquet model does not have time-reversal invariance and has CUE level statistics in its thermal phase, while the Hamiltonian model does have time-reversal invariance and hence GOE statistics, thus the difference in the thermal values of $\langle r\rangle_\text{CUE}\approx0.5996$ vs. $\langle r\rangle_\text{GOE}\approx 0.5307$ \cite{atas_distribution_2013}.
	 \label{fig:mean_r}}
\end{figure}
 
\subsection{Model characterization}

In order to get oriented with our new Floquet model and compare it to the more familiar Hamiltonian model, in Fig.~\ref{fig:mean_r} we show a common diagnostic that has been extensively studied in the context of MBL: the mean level spacing ratio $\langle r \rangle$~\cite{Oganesyan-Huse2007}. 
This is a dimensionless quantity which measures the average level repulsion in the spectrum, and the location of the finite-size crossing in this quantity marks the crossover between the thermal regime and the finite-size MBL regime, which we denote by $\mathcal{L}_r$. 

As we will argue below, for the sizes accessible to numerics, this crossover is {\it not} a good estimate of where the thermal-to-MBL {\it phase transition} is in the limit of large $L$. Instead, the lower-bound on the location of the phase transition, that we estimate by testing for stability to avalanches, is at a much larger disorder strength than $\mathcal{L}_r$ for accessible $L$. Since we argue that this feature in $\langle r\rangle$ is not a relevant estimate of the MBL phase transition but instead is a measure of the finite-size MBL to thermal crossover, we could just as reasonably have used, say, the midpoints of the changes in $\langle r\rangle$ from its random matrix theory (RMT) value to its Poisson value instead of the crossings.  But, to keep more contact with previous work, for now we will stick with using the crossings to define this landmark.

The mean level spacing ratio is computed by averaging $r_n = \min(\delta_n, \delta_{n-1}) / \max(\delta_n, \delta_{n-1})$ over eigenstates and realizations, where $\delta_n$ is the magnitude of the spectral gap between (ordered) eigenvalues $n$ and $n+1$. 
For the Floquet model, eigenvalues are naturally ordered on the unit circle by increasing phase. For system sizes $L\leq 14$, we use all eigenvalues of $U$ obtained using exact diagonalization, and a number of disorder realizations which varies in the range $10^4 - 4\cdot 10^4$. For $L\geq 16 $, we use the $50$ eigenvalues closest to $1$, calculated using geometric sum filtering \cite{Luitz2021} for $3000-6000$ realizations. 
For the Hamiltonian model we average over the middle fifth of states in the spectrum and $8000-64,000$ disorder realizations for $L\leq 16$ . For larger sizes, we take advantage of the sparseness of $H$ and use shift-invert diagonalization \cite{luitz_many-body_2015,pietracaprina_shift-invert_2018} to obtain the central 50 eigenvalues for $500-8000$ disorder realizations. We exclude eigenvalues further away from the center of the spectrum to avoid the most significant effects of the energy dependence of the eigenstates.  

\subsection{Floquet random circuit coupled to an infinite bath \label{sec:floquet_superop}}

Our unitary Floquet model is described by the Floquet operator $U$. It corresponds to the action of the unitary superoperator $\mathcal{S}_U[\rho] = U \rho U^\dagger$. We now formally extend the spin chain by one extra spin on site $0$, in contact with the first site $i=1$. This spin acts as the rightmost spin of a coupled bath.
In each cycle, we reset the state of this spin to a $2\times 2$ featureless density matrix $\frac{1}{2} \mathbb{1}$. 
The coupling to the rest of the chain is given by $u_0 = \exp( \mathrm{i} M_0 / \alpha)$, sampled from the same distribution as the other two-site gates. 
Tracing out the bath spin $0$ at the end of the cycle, the action of the superoperator can be expressed diagrammatically as
\begin{equation}
    \mathcal{S}[\rho] = 
    \begin{tikzpicture}[baseline=(current bounding box.center), scale=1]
        \draw[thick, fill=black!10!white,rounded corners=2pt] (0,-0.30) rectangle (2,0.3); 
        \draw[thick, fill=Ured,rounded corners=2pt] (0,0.50) rectangle (2,1.1); 
        \draw[thick, fill=Ublue,rounded corners=2pt] (0,-0.50) rectangle (2,-1.1); 

        \draw[thick, fill=black!10!white,rounded corners=2pt] (-0.8,-0.30) rectangle (-0.2,0.3); 
        \draw[thick, fill=Ured,rounded corners=2pt] (-0.8,1.30) rectangle (0.4,1.9); 
        \draw[thick, fill=Ublue,rounded corners=2pt] (-0.8,-1.30) rectangle (0.4,-1.9); 

        \draw (1,0.0) node {$\rho$}; 
        \draw (1,0.8) node {$U$}; 
        \draw (1,-0.8) node {$U^\dagger$}; 
        \draw (-0.5,0.0) node {$\frac 1 2 \mathbb{1}$}; 
        \draw (-0.2,1.6) node {$u_{0}$}; 
        \draw (-0.2,-1.6) node {$u_{0}^\dagger$}; 

        \draw[thick] (-0.5, 0.3) -- (-0.5, 1.3);
        \draw[thick] (-0.5, -0.3) -- (-0.5, -1.3);
        \draw[thick] (-0.5,1.9) to [out=120, in=-120] (-0.5,-1.9);

        \draw[thick] (0.2, 0.3) -- (0.2, 0.5);
        \draw[thick] (0.2, 1.1) -- (0.2, 1.3);
        \draw[thick] (0.2, 1.9) -- (0.2, 2.0);

        \draw[thick] (0.2, -0.3) -- (0.2, -0.5);
        \draw[thick] (0.2, -1.1) -- (0.2, -1.3);
        \draw[thick] (0.2, -1.9) -- (0.2, -2.0);

        \foreach \x in {0.6,1,1.4,1.8}
        {
            \draw[thick] (\x, 1.1) -- (\x, 1.3);
            \draw[thick] (\x, 0.3) -- (\x, 0.5);
            \draw[thick] (\x, -0.3) -- (\x, -0.5);
            \draw[thick] (\x, -1.1) -- (\x, -1.3);
        }
    \end{tikzpicture}
    =
    \begin{tikzpicture}[baseline=(current bounding box.center), scale=1]
        \draw[thick, fill=black!10!white,rounded corners=2pt] (0,-0.30) rectangle (2,0.3); 
        \draw[thick, fill=Ured,rounded corners=2pt] (0,0.50) rectangle (2,1.1); 
        \draw[thick, fill=Ublue,rounded corners=2pt] (0,-0.50) rectangle (2,-1.1); 

        \draw[thick, fill=Ugreen] (0.4,1.2) -- (0.4,1.5) -- (-0.5,1.5) -- (-0.5, -1.5) -- (0.4,-1.5) -- (0.4, -1.2) -- (-0.1, -1.2) -- (-0.1,1.2) -- cycle ;

        \draw (1,0.0) node {$\rho$}; 
        \draw (1,0.8) node {$U$}; 
        \draw (1,-0.8) node {$U^\dagger$}; 
        \draw (-0.3,0.) node {$D$}; 
       
        \draw[thick] (0.2, 1.1) -- (0.2, 1.2);
        \draw[thick] (0.2, 0.3) -- (0.2, 0.5);
        \draw[thick] (0.2, -1.1) -- (0.2, -1.2);
        \draw[thick] (0.2, -0.3) -- (0.2, -0.5);
        \draw[thick] (0.2, 1.5) -- (0.2, 1.6);
        \draw[thick] (0.2, -1.5) -- (0.2, -1.6);
        
        \foreach \x in {0.6,1,1.4,1.8}
        {
            \draw[thick] (\x, 1.1) -- (\x, 1.3);
            \draw[thick] (\x, 0.3) -- (\x, 0.5);
            \draw[thick] (\x, -0.3) -- (\x, -0.5);
            \draw[thick] (\x, -1.1) -- (\x, -1.3);
        }
    \end{tikzpicture}.
    \label{eq:Floquet_superoperator}
\end{equation}
In the last step, we have introduced the tensor $D$, given by:
\begin{equation}
2 D_{i_1, i_1'}^{j_1, j_1'} = \sum_{i_0, j_0, i_0', j_0'}   u_0^{(i_0, i_1), (i_0', i_1')} {u_0^\dagger}^{(j_0', j_1'),(j_0, j_1)} \delta_{i_0, j_0} \delta_{i_0', j_0'}.
    \label{eq:Dgate}
\end{equation}

\begin{figure}
    \centering
    \includegraphics[width=\columnwidth]{{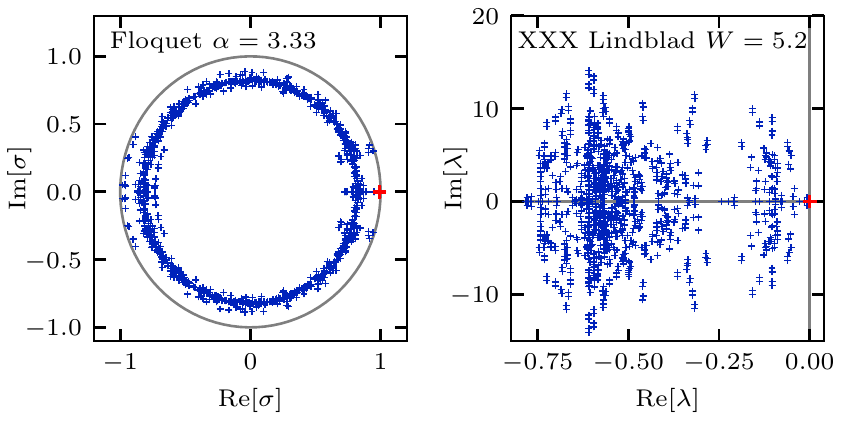}}
    \caption{Example spectra of the two superoperators for $L=5$ spins. \textbf{Left:} Floquet superoperator with $\alpha=3.33$. The spectrum (blue) is strictly contained inside the unit circle (gray). The slowest mode (red) is the eigenvalue with the largest modulus less than $1$. \textbf{Right:} Liouvillian of the random field Heisenberg  model coupled to a bath. The spectrum lies in the complex left half plane with nonpositive real parts. The slowest mode (red) is the eigenvalue whose real part is negative and is closest to zero.  }
    \label{fig:spectra}
\end{figure}

This Floquet superoperator $\mathcal{S}$ (Eq.~\ref{eq:Floquet_superoperator}) describes the stroboscopic dynamics of the density matrix of the system:
\begin{equation}
    \rho(t+1) = \mathcal{S}[\rho(t)] = \mathcal{S}[\mathcal{S}[\rho(t-1)]] = \mathcal{S}^{t+1}[\rho(t=0)].
    \label{eq:superop_dynamics}
\end{equation}
$\mathcal{S}$ is represented by a subunitary matrix in operator space, i.e., its spectrum $\{ \sigma_n\}$ is contained inside the complex unit disk, $|\sigma_n|\leq 1$, (cf. Fig. \ref{fig:spectra}). While one eigenvalue $\sigma_0=1$ exists, it corresponds to the steady state $R_0=\frac 1 Z \mathbb 1$, since $S[R_0] = R_0$, as can be seen from Eq.~\ref{eq:Floquet_superoperator} due to the unitarity of $U$ and $u_0$.

We can use the left and right eigenmatrices $L_i$, $R_i$ of $\mathcal S$ with $\mathcal{S} R_i = \sigma_i R_i$, $L_i \mathcal{S} = \sigma_i L_i$ and $\Tr(L_i^\dagger R_j) = \delta_{ij}$ to calculate the time evolution of any initial state $\rho(t=0)$:
\begin{equation}
    \rho(t) = R_0 + \sigma_1^t c_1 R_1 + \sum_{k>2} \sigma_k^t c_k R_k, 
\end{equation}
with $c_k=\Tr(L_k \rho(t=0))$. Here, we order the eigenvalues $\sigma_i$ by their modulus, $|\sigma_i|\geq|\sigma_j|$ if $j\geq i$, such that $\sigma_0=1$ corresponds to the steady state $R_0=\frac 1 Z \mathbb 1$, and $\sigma_1$ is the eigenvalue with the second largest modulus, and thus represents the slowest relaxation rate $1/\tau = -\ln |\sigma_1| $ in the system. 
The corresponding mode $R_1$ decays as 
\begin{equation}
    \sigma_1^t = \exp( t \ln |\sigma_1| ), 
    \label{eq:slowmode}
\end{equation}
and therefore $\tau = -1/\ln |\sigma_1|$  is the longest timescale in the system. In an MBL system, this timescale is determined by the couplings between the bath and the farthest l-bit at the other end of the chain.
We calculate the spectral gap using (dense) shift-invert diagonalization of the superoperator $\mathcal{S}$, targeting eigenvalues closest to $1$ (euclidian distance  in the complex plane), and we have checked that this captures the slowest decay rate.

\subsection{Hamiltonian coupled to an infinite bath }
\label{sec:hamiltonian_superop}

In order to study thermalization rates in the Hamiltonian model we introduce a coupling to an infinite bath to the spin located at the left edge of the chain. The dynamics of this system is described by the master equation:

\begin{eqnarray}
\label{eq:time_evo_rho}
 \frac{d\rho(t)}{d t}  = \mathcal{L}[\rho],
\end{eqnarray}

\begin{eqnarray}
\label{eq:lindbladian}
 \mathcal{L}[\rho] = -i[H,\rho]+\sum_{\mu\nu}K_{\mu\nu}\left( L_{\mu}\rho L^{\dagger}_{\nu}-\frac{1}{2}\{L^{\dagger}_{\nu}L_{\mu},\rho\} \right),
\end{eqnarray}
where the Lindblad operators $L_{\mu}=( X_1,Y_1,Z_1 )$ are the Pauli operators acting on the left-most spin. The Lindblad coupling breaks the $U(1)$ symmetry of the XXX Hamiltonian so all magnetization sectors are mixed and the full operator Hilbert space dimension is $4^L$.   The eigenvalues $D$ of the Kosakowski matrix $K$ are sampled from a uniform distribution and normalized such that $\Tr D=2$. From $D$, we obtain the Kosakowski matrix $K=U^{\dagger}DU$ where $U$ is a random matrix from the circular unitary ensemble (CUE), similarly to the sampling in Ref. \cite{wang_hierarchy_2020}. As in the Floquet case in Sec. \ref{sec:floquet_superop}, the solution of Eq. \eqref{eq:time_evo_rho} is obtained from the eigenmodes of the Lindblad superoperator:
\begin{eqnarray}
\label{eq:time_solution}
 \rho(t) = R_0+ e^{\lambda_1 t}c_1 R_1+\sum_{k>2} e^{\lambda_k t}c_k R_k,
\end{eqnarray}
where $R_k$ are the right eigenmatrices of $\mathcal{L}$, and $c_k=\Tr(L_k \rho(t=0))$ are the overlaps of the initial state with the left eigenmatrices of $\mathcal{L}$. Left and right eigenmatrices are orthogonal to each other such that $\Tr(L_iR_j)=\delta_{ij}$.
The eigenvalue $\lambda_0=0$ corresponds to the steady state $R_0=\frac{1}{Z} \mathbb{1}$, 
and the eigenvalue $\lambda_1$ with the second largest real part $\mathrm{Re}(\lambda_1)<0$ 
represents the slowest decay mode in the system.
We identify $\tau=-1/\mathrm{Re}{(\lambda_1})$ as the time scale on which the entire system, including the farthest l-bit, reaches the steady state. 
As in the Floquet case, we use shift-invert diagonalization to obtain the slowest decay rate of the system, exploiting the fact that $\mathcal L$ is a sparse matrix.

\subsection{Weak bath-coupling limit of the Floquet circuit\label{sec:weak_coupling_floquet}}

In work following our first manuscript version of this paper~\cite{Morningstar-Huse2021}, Sels showed that the limit of weak coupling to the infinite bath is sufficient for studying the slowest rate of thermalization, and allows the numerical calculations to reach larger system size $L$~\cite{sels_markovian_2021}. We therefore introduce versions of our open Floquet and Hamiltonian models (described in Secs. \ref{sec:floquet_superop} and \ref{sec:hamiltonian_superop}) in this weak coupling limit. This allows us to extend our analysis of the landmark $\mathcal{L}_\text{avch}$ to larger system sizes and disorder than what was done in our initial version.

In Eq.~\ref{eq:Floquet_superoperator} the ``ancilla qubit", which acts as the rightmost spin of the bath, is in a maximally mixed state. The leftmost spin of the chain is coupled through all channels with this bath spin, so the dissipation is maximal. In order to be able to tune the level of dissipation, we replace the $D$ superoperator by:
\begin{eqnarray}
\label{eq:d_gate_weak_coupling}
 D[\rho] = \dfrac{\rho}{1+3\gamma}+\dfrac{\gamma}{1+3\gamma}\sum_{\mu} E_\mu\rho E^{\dagger}_\mu,
\end{eqnarray}
where $E_{\mu} = (X_1,Y_1,Z_1)$ are Krauss operators on the first spin and $\gamma$ is the parameter that tunes the dissipation strength. The limit $\gamma\to 0$ recovers unitary dynamics, whilst $\gamma\rightarrow\infty$ is the maximal dissipation limit similar to the $D$ gate shown in Eq. \ref{eq:Dgate}, with the difference being that dissipation is now homogeneous on all allowed channels. Eq.~\ref{eq:d_gate_weak_coupling} allows us to study the weak coupling limit ($\gamma \ll 1$) of the Floquet dissipative circuit in a controlled manner. We continue to denote the Floquet superoperator for one period of evolution by $\mathcal{S}[\rho] = D[U\rho U^\dagger]$.
 
In the unitary limit $\gamma=0$, the eigenvalues and eigenoperators of the superoperator $\mathcal{S}$ are products of eigenvalues and eigenstates of the Floquet unitary: $\sigma_{nm} = e^{i(\theta_n-\theta_m)}$ and $\rho_{nm} = |n\rangle\langle m|$, where $U|n\rangle = e^{i\theta_n} |n \rangle $.
Thus there are $2^L$ degenerate operators with unit eigenvalue, corresponding to $n=m$.
In the limit of nonzero but small $\gamma$, the dissipation acts as perturbation of the unitary evolution, allowing a perturbative treatment in the basis of eigenstates $|n\rangle\langle n|$. 
In this subspace, the matrix elements of the Floquet superoperator are
\begin{eqnarray}
\mathcal{S}_{nm} &=& \langle m| D\left[|n\rangle\langle n|\right] |m \rangle \nonumber\\
&=& \dfrac{ \delta_{nm}}{1+3\gamma}+\dfrac{\gamma}{1+3\gamma}\sum_{\mu} \langle m | E_\mu|n\rangle \langle n |E^{\dagger}_\mu |m\rangle.
\end{eqnarray}
By diagonalizing this matrix we obtain a perturbative estimate of the slowest mode and associated rate of thermalization in the dissipative Floquet dynamics. Note that in this perturbative treatment, we build and diagonalize a matrix of linear size $2^L$, not $4^L$ as is done when working nonperturbatively. We have set $\gamma=0.001$ throughout the entire text when dealing with perturbative dissipation. There are additional issues related to numerical precision explained in detail in Appendix~\ref{app:quad-precision}.

\subsection{Weak bath-coupling limit of the Hamiltonian\label{sec:weak_coupling_hamiltonian}}

Again, in order to simplify the study of the thermalization rate of a spin chain coupled to an infinite bath at one end, in the bath-coupled Hamiltonian system the weak coupling limit is considered, similar to what was done in Sels' follow up to our original work~\cite{sels_markovian_2021}. 

In the dissipationless limit (Lindblad superoperator with Kosakowski matrix set to zero) the eigenvalues of the Lindbladian are $\lambda=i(E_n-E_m)$ with the set of eigenoperators $\rho=|n\rangle\langle m|$, where $|n\rangle$ are the eigenstates of the Hamiltonian and $E_n$ their corresponding eigenvalues. There are $2^L$ zero eigenvalues and the rest fall on the imaginary axis and come in conjugate pairs. When the dissipation is perturbative, the slowest mode is well approximated within the degenerate subspace of operators $|n\rangle\langle n|$~\cite{sels_markovian_2021}.
Starting from Eq.~\ref{eq:lindbladian}, the Kosakowski matrix is now diagonal, $K_{\mu\nu}=\gamma \delta_{\mu\nu} $ with $\gamma=0.001 $, and the jump operators remain the same. The matrix elements of the Linbladian in the degenerate sector read
\begin{eqnarray}
\mathcal{L}_{nm} & = & \langle m | \mathcal{L}\big[|n\rangle\langle n|\big] |m\rangle \nonumber \\
& = & \gamma \sum_{\mu} \Big[\langle m | L^{\dagger}_\mu|n\rangle \langle n |L_\mu |m\rangle-\delta_{nm}\langle m | L^{\dagger}_\mu L_\mu|n\rangle\Big] \nonumber \\ 
& = & -3\gamma\delta_{nm}+\gamma \sum_{\mu}|\langle m | L_\mu|n\rangle |^2.
\end{eqnarray}
Similar to the Floquet case, the perturbative approximation reduces the problem of finding the slowest mode to diagonalizing a dense matrix of size $2^L\times 2^L$, rather than diagonalizing the full $4^L\times 4^L$ superoperator matrix. Choosing a diagonal Kosakowski matrix ensures that the resulting matrix $\mathcal{L}_{nm}$ is Hermitian. We checked that relaxing that condition does not change the qualitative behavior of the slowest mode.  Constructing $\mathcal{L}_{nm}$ requires all Hamiltonian eigenstates in all magnetization sectors, and the diagonalization of each sector is carried out separately. The numerical bottleneck is diagonalizing the dense Lindbladian matrix of size $2^L$. Another issue is the insufficiency of double precision arithmetic, which is the case for large $L$ and strong disorder in both Hamiltonian and Floquet models. This issue is further addressed in Appendix \ref{app:quad-precision}.

\section{Open System Results\label{sec:results_open}}

We begin the discussion of our results by considering the avalanche instability of MBL chains using the two dissipative models introduced in Secs.~\ref{sec:floquet_superop} and \ref{sec:hamiltonian_superop}.

As outlined before, we focus on the spectral gap of the superoperators $\mathcal{S}$ and $\mathcal{L}$ describing MBL chains coupled to an infinite bath at one end, since it encodes the slowest decay rate $1/\tau$ towards the steady state.
The coupling to an infinite bath ensures that the system evolves towards an infinite temperature steady state $\rho_0 = \frac{1}{Z} \mathbb{1}$ throughout the entire phase diagram, i.e., for any strength of disorder.
We interpret the slowest decay rate as the thermalization rate of the chain, and how this quantity scales with $L$ determines if the associated isolated chain is unstable to avalanches or not. Exemplary spectra of the two superoperators are shown in Fig.~\ref{fig:spectra}, illustrating that the spectrum of the Floquet superoperator $\mathcal{S}$ is contained inside the unit disk, while the spectrum of the Liouvillian $\mathcal{L}$ resides in the left half plane of the complex plane, since all components not in the steady state vanish at long times. 
The eigenvalue of $\mathcal{S}$ ($\mathcal{L}$) with the second largest modulus (real part) $\sigma_1$ ($\lambda_1$) encodes the slowest timescale $\tau$ of decay to the steady state $\rho_0$ corresponding to $\sigma_0=1$ ($\lambda_0=0$). 

\subsection{Slowest decay rate in the presence of an infinite bath}

\begin{figure}
\includegraphics[width=1.0\linewidth]{{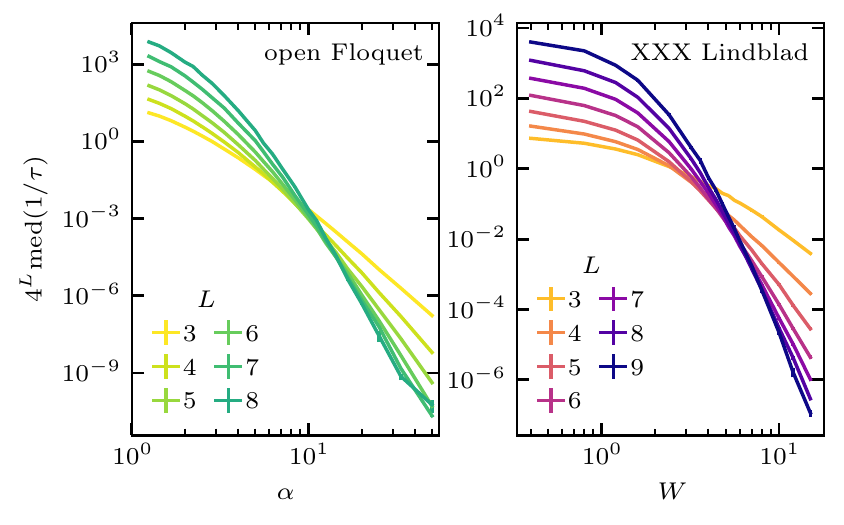}}
\caption{Typical slowest decay rate as a function of disorder strength. The curves are the (scaled) median of the distribution of $1/\tau$ over realizations for the Floquet and Lindblad super-operators. The crossings occur when the scaling of the decay rates with system size is $1/\tau\propto 4^{-L}$ and thus indicate the location of $\mathcal{L}_\text{avch}$. Smaller sizes $L<7$ are calculated using full diagonalization, bigger sizes are computed using shift-invert diagonalization with target eigenvalue $\lambda=0$ (Lindblad) and $\sigma=1$ (Floquet). For the open Floquet system, we used at least $20000, 20000, 10000, 4000, 1000, 500$ disorder realizations for $L=3,4,5,6,7,8$ respectively. For the Lindblad operator, we collected
$5000, 10000, 10000, 8000, 9000, 8000, 1000$ disorder realizations for $L=3,4,5,6,7,8,9$.
Error  bars  are  68\%  bootstrap  confidence intervals. 
\label{fig:crossing_slowest_rate}}
\end{figure}

In Fig. \ref{fig:crossing_slowest_rate} we analyze the scaling, with system size $L$, of the typical slowest decay rate $1/\tau$ as a function of the strength of disorder ($\alpha$ in the Floquet case and $W$ in the Hamiltonian model). At small disorder, deep in the thermal phase, the slowest rate of decay scales as a power of $L$, and is determined by the speed of information scrambling in the Floquet case ($1/\tau \propto L^{-1}$), and by hydrodynamic modes in the Hamiltonian case ($1/\tau \propto L^{-2}$). 
On the other hand, in the MBL phase, the typical decay rate towards the thermal steady state is exponentially small in $L$, $\propto k^{-L}$. As explained in Sec.~\ref{sec:avalanche_intro}, the avalanche instability occurs if the slowest decay rate scales with $L$ more slowly than $1/\tau \propto 4^{-L}$. Thus the product of the typical decay rate and the scaling factor $4^L$ of the avalanche instability increases with $L$ in the thermal phase and decreases with $L$ in the MBL phase. Fig.~\ref{fig:crossing_slowest_rate} shows this change of behavior as a function of the disorder parameter $\alpha$ ($W$) in the open Floquet (Hamiltonian) model.

Ideally one would perform a scaling collapse of these curves to estimate the location of the avalanche-driven phase transition. However, the appropriate form of the scaling function one should use is not clear; recent RG approaches predict a two-parameter scaling theory similar in some respects to Kosterlitz-Thouless scaling, but we know that the small system sizes accessible to numerics are far from the scaling regime controlled by the asymptotic fixed point. 
Therefore, we simply identify the location of the finite-size crossing of the curves in Fig.~\ref{fig:crossing_slowest_rate}, for the largest systems we can access, as a lower bound on $\mathcal{L}_\text{avch}$ in the limit of $L\to\infty$, assuming a monotonic drift with $L$. Strikingly, even at the small system sizes accessible to our open system calculations, $\mathcal{L}_\text{avch}$ occurs at much stronger disorder strengths compared to the reference $\mathcal{L}_r$ shown in Fig. \ref{fig:mean_r}: roughly at $\alpha>13$ in the Floquet case, and $W>7$ in the open Hamiltonian model. 

\begin{figure}
\includegraphics[width=1.0\linewidth]{{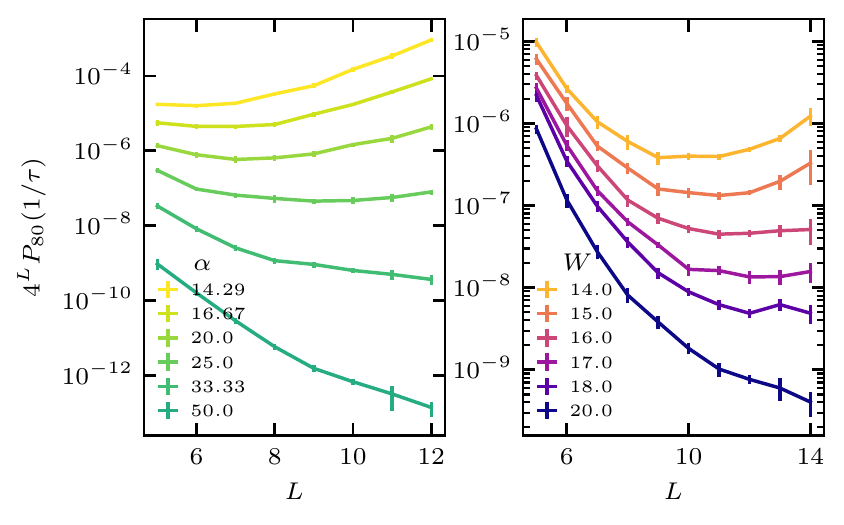}}
\caption{Typical decay rate as a function of system size computed using perturbation theory in the weak bath-coupling limit. The curves are the (scaled) $80^\mathrm{th}$ percentile of the distribution of $1/\tau$ over realizations for the Floquet and Lindblad super-operators. System sizes are $L=5-12$ and $L=5-14$ for Floquet and Lindblad set ups, respectively. At least 5000 disorder realizations were used for each set of parameters except for those computed using quadruple precision and $L=14$ (Lindblad) or $L=12$ (Floquet), for which 1500-2000 realizations were used. Error  bars  are  68\%  bootstrap  confidence intervals.}
\label{fig:crossing_slowest_rate_pt_system_size}
\end{figure}

In order to access larger system sizes we study the weak bath-coupling limit that allows a perturbative treatment (see Sec. \ref{sec:weak_coupling_floquet} and \ref{sec:weak_coupling_hamiltonian} for more details). The results are presented in Fig. \ref{fig:crossing_slowest_rate_pt_system_size}. We look at the $80^\mathrm{th}$ percentile, as was done in Ref.~\cite{sels_markovian_2021}, rather than the median ($50^\mathrm{th}$ percentile) because it helps to mitigate issues with numerical precision present at stronger disorders and larger system sizes (further discussion of the numerical issue is found in Appendix~\ref{app:quad-precision}). In the Floquet model the typical rate $1/\tau$ decays faster than $4^{-L}$ at accessible $L$ only for $\alpha = 33.33$ and $\alpha= 50$, while the curves at smaller $\alpha$ eventually scale slower than $4^{-L}$ at the largest system sizes that we have data for. In the Lindblad model, the decay rate scales slower than $4^{-L}$ at disorders $W=14.0,15.0$ and the largest $L$, and faster only at $W=20.0$, with disorders $W=16.0,17.0,18.0$ showing plateaus (within error bars) that indicate an effective critical region for the present system sizes. Based on this perturbative analysis, the landmark $\mathcal{L}_\text{avch}$ is pushed even farther away from the standard landmark $\mathcal{L}_r$, roughly at $\alpha>25$ and $W>18$ for the Floquet and Lindblad model respectively. The resulting wide range of disorder in between these two landmarks is thus part of the finite-size MBL regime of the thermal phase that we are exploring in these finite-size systems.

\subsection{Distributions of the slowest decay rates}

\begin{figure}
\includegraphics[width=1.0\linewidth]{{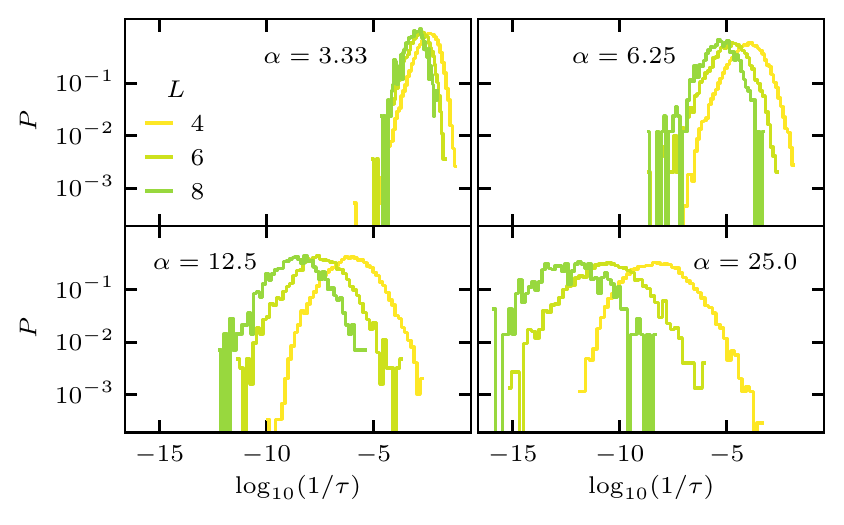}}
\caption{Distributions over realizations of the slowest decay rate $1/\tau$ of the Floquet super-operator for disorder strengths $\alpha=3.33,6.25,12.5,25.0$. This is part of the data shown in Fig.~\ref{fig:crossing_slowest_rate}. \label{fig:distribution_decay_rate_floquet}}
\end{figure}

The distribution of the slowest decay rates of the Floquet superoperator are shown in Fig. \ref{fig:distribution_decay_rate_floquet} (cf. Appendix~\ref{app:superop-dist} for data for the Lindblad superoperator) and are approximately log-normal.  The variance of the logarithm of the rate is consistent with being $L$-independent at the larger values of $L$. The peaks of the distributions scale with the system size, and the decrease of the mode of the distribution for larger system sizes $L$ at large disorder $\alpha$ is reflected in the decrease of the median of the distribution shown in Fig. \ref{fig:crossing_slowest_rate}.

\section{Undoing many-body resonances\label{sec:untangle}}

Before moving on to study many-body resonances in eigenstates in the MBL regime in Sec.~\ref{sec:results}, in this section we introduce a tool for studying such resonances. It will allow us to associate an off-diagonal matrix element, to any two eigenstates, that is responsible for the level repulsion (however strong or weak) between those two levels, and some of the entanglement in those eigenstates.

Many-body resonances between configurations of local degrees of freedom manifest themselves as entanglement in the eigenstates of the dynamics. 
This entanglement arises during a basis rotation from a localized basis, in which states are dynamically connected by off-diagonal matrix elements of the Floquet unitary (or Hamiltonian), to a basis of entangled eigenstates, which are not dynamically connected. This idea is realized explicitly in Wegner-Wilson flows~\cite{Pekker-Refael2017}, but that is not what we do below. Instead, we consider a hypothetical flow, which at its end arrives at the eigenstates of the dynamics. We are essentially interested in the very last steps of this flow, which rotate away the last off diagonal elements of $U$ (or $H$).
Depending on the location along this hypothetical flow, resonances can be indicated by entanglement in the set of states that is flowing, or by nonzero off-diagonal matrix elements that couple the states. This description is qualitative, and in this section we aim to introduce a quantitative procedure for moving between these two views in a controlled setting.

We want the ability to ``undo" some of the entanglement associated with many-body resonances that exists in the basis of eigenstates, and transform to a different, more localized basis in order to study the underlying matrix elements that are responsible for that entanglement, ``rewinding'' the hypothetical flow by the last steps. 
It is unclear to us how to do this meaningfully when there are many states involved, so in this section we describe a method for doing this transformation explicitly with two states treated in isolation. In other words, we develop a procedure for transforming any two eigenstates into two more-localized states that they are superpositions of. Then, since those two more-localized states are not eigenstates, they do have a nonzero off-diagonal matrix element that connects them in the dynamics, and we study these matrix elements in Sec.~\ref{sec:results}.

\begin{figure}
\centering
\includegraphics[width=0.9\linewidth]{{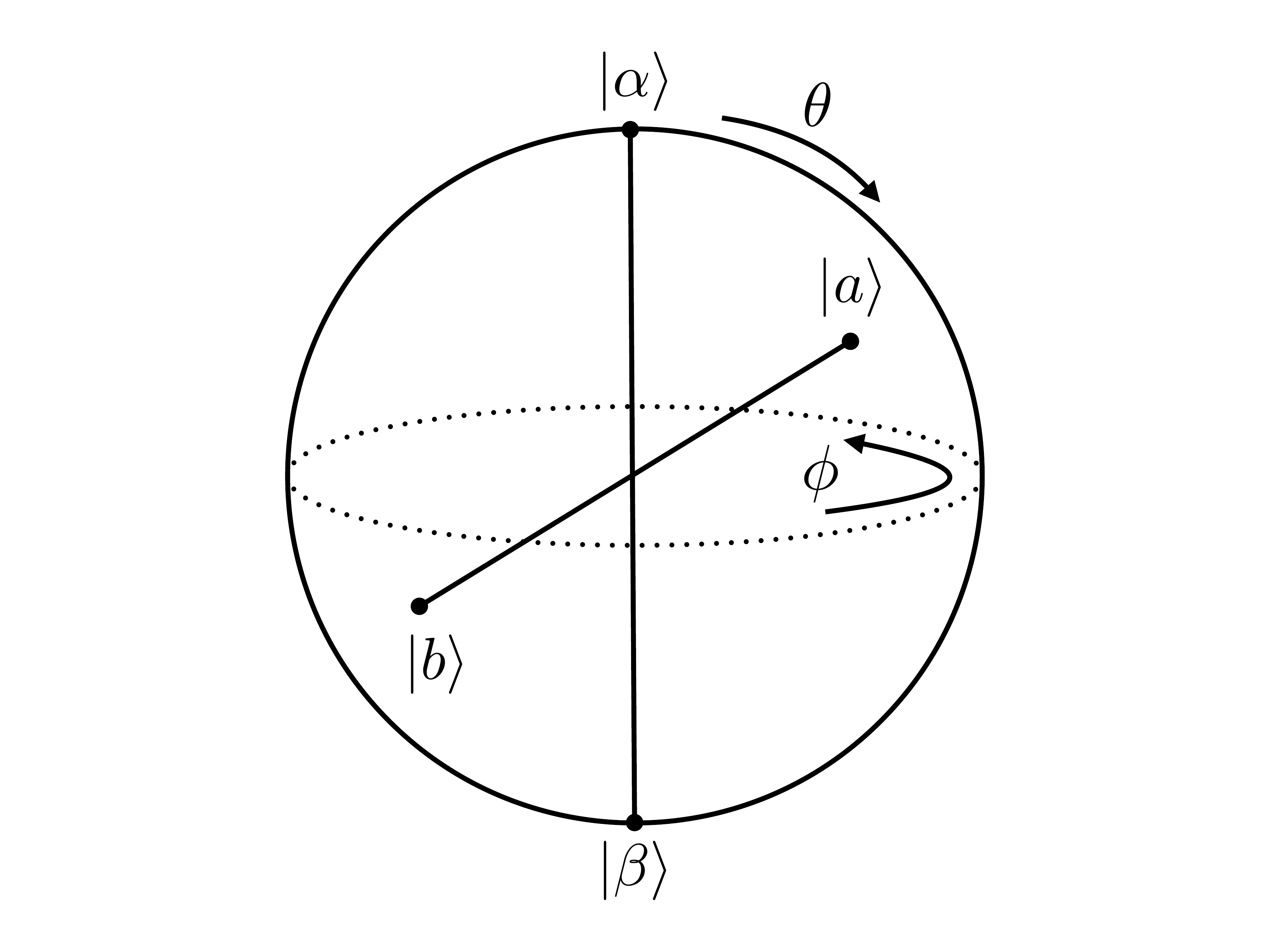}}
\caption{The Bloch sphere representing the 2D subspace spanned by two eigenstates, and the rotation to the corresponding two demixed states. A Bloch sphere can be defined using any two orthogonal states. Here we associate the poles to the two eigenstates $|\alpha\rangle$ and $|\beta\rangle$, and the points that represent the demixed states $|a\rangle$ and $|b\rangle$ are rotated away from the poles at polar and azimuthal angles $\theta$ and $\phi$. Note that orthogonal states in the 2D subspace represented by this Bloch sphere are antipodal on the surface of the sphere. \label{fig:bloch_sphere}}
\end{figure}

Consider two eigenstates of the dynamics, $| \alpha \rangle$ and $| \beta \rangle$. These can be eigenstates of a Floquet operator or a Hamiltonian, but in this section we will consider a Floquet system for concreteness. We want to find the two states
\begin{eqnarray}
    &&|a\rangle = \cos \left(\frac{\theta}{2}\right) |\alpha\rangle + \text{e}^{i\phi} \sin \left(\frac{\theta}{2}\right) |\beta\rangle \label{eq:a_state} \\
    &&|b\rangle = - \text{e}^{-i\phi} \sin \left(\frac{\theta}{2}\right) |\alpha\rangle + \cos \left(\frac{\theta}{2}\right) |\beta\rangle \label{eq:b_state}
\end{eqnarray}
that are the orthogonal superpositions of $| \alpha \rangle$ and $| \beta \rangle$ that are as localized as possible. We refer to $|a\rangle$ and $|b\rangle$ as the ``demixed" states. The rotation from the two eigenstates to the more localized, demixed states is parametrized by two angles, $\theta \in [0,\pi]$ and $\phi\in[0,2\pi]$, on a Bloch sphere whose poles are defined by the two eigenstates (see Fig.~\ref{fig:bloch_sphere}). We emphasize that a Bloch sphere can be constructed from any two orthogonal states, and here we are defining the two eigenstates to be at the poles, while the demixed states are rotated away from the poles.

As mentioned earlier, we can imagine the demixing procedure as a reverse renormalization group flow in this two-dimensional (2D) subspace that starts at the eigenstates and moves towards the localized $Z$ basis states as much as possible, ending at the demixed states. During this process the Floquet unitary matrix, when expressed in the 2D basis $\{ | a \rangle, | b \rangle \}$, goes from being diagonal when $\theta = 0$ to having a nonzero off-diagonal matrix element $U_{ab}$ (and its conjugate) which couples the states $| a \rangle$ and $| b \rangle$ under the dynamics. Note that the matrix elements of the Floquet unitary between the demixed states and any other eigenstate outside of this 2D subspace are still zero, so only one isolated nonzero off-diagonal matrix element is generated by rotating two of the eigenstates into a superposition. In the case that the two eigenstates are a well-isolated, strong, two-state resonance, $| a \rangle$ and $| b \rangle$ will both have an $O(1)$ overlap with both $| \alpha \rangle$ and $| \beta \rangle$, i.e., $\theta$ will be comparable to $\pi/2$ in Eqs.~\ref{eq:a_state} and \ref{eq:b_state}.

\begin{figure}
\includegraphics[width=0.9\linewidth]{{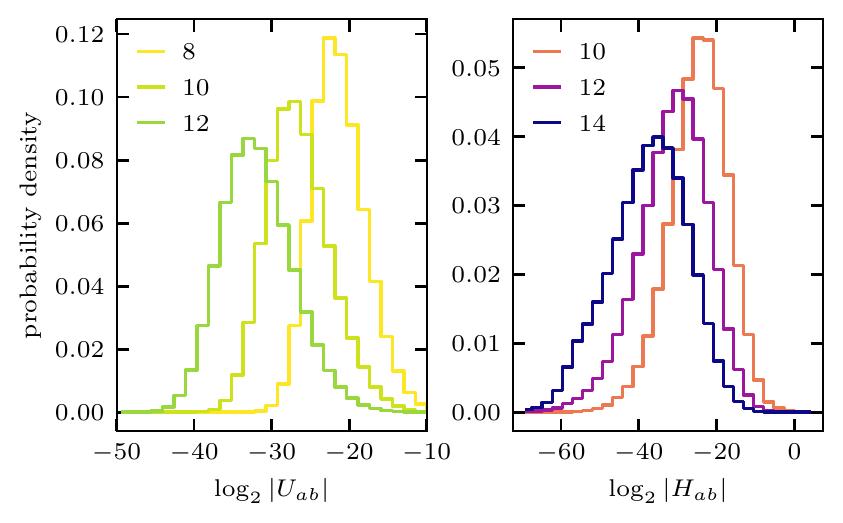}}
\caption{Distributions of matrix elements for end-to-end processes in the MBL regime. These distributions are collected over such pairs of adjacent eigenstates (see main text), and extracted via the method detailed in Sec.~\ref{sec:untangle}. \textbf{Left:} The Floquet circuit model. The data was collected from $10^4, 10^4, 1.5\times 10^3$ realizations at $\alpha = 10$ and $L=8, 10, 12$, respectively. \textbf{Right:} The Hamiltonian model. The data was collected from $10^4$ realizations at $W=6$ and $L=10,12,14$. See Appendix~\ref{app:num_precision} for a discussion of numerical errors in finite-precision arithmetic. \label{fig:Uab_Hab_dist}}
\end{figure}

In practice, the way we find an appropriate rotation is to maximize 
\begin{eqnarray}
    f(\theta, \phi) = \sum_{i=1}^L Z_{\alpha\alpha} Z_{aa} + Z_{\beta\beta} Z_{bb} \label{eq:maximize} 
\end{eqnarray}
over the angles $\theta$ and $\phi$, where $i$ runs over sites, we have suppressed the site index $i$ on the Pauli $Z$ operators, and $Z_{\psi \varphi}$ is the matrix element $\langle\psi|Z_i|\varphi\rangle$. We have chosen to maximize this particular function because it approximates the sum of squared $Z$ magnetizations over all sites and both demixed states, $F(\theta, \phi) = \sum_{i=1}^L Z_{aa}^2 + Z_{bb}^2 $, which is maximized by the computational basis states, and because the maximum of $f$ can be found analytically given data on the matrix elements of $Z_i$ in the basis of eigenstates. In considering the approximation of substituting $f$ for $F$ we should consider what we need out of this approximation. As we will see later, the distribution of matrix elements $|U_{ab}|$ over pairs of states is very broad on a log scale (see Fig.~\ref{fig:Uab_Hab_dist}), and this implies the distribution of $\theta$ is also broad (see Eq.~\ref{eq:Uab}). So, as long as we can determine $\theta$ accurately on a log scale, that is sufficient for our study ($\phi$ is not of much importance). Indeed when the maximum of $F$ is at $\theta \ll 1$, the maximum of $f$ will approximate this very well, and when the maximum of $F$ is at $\theta \sim 1$, then all we need is that the maximum of $f$ is at $\theta\sim 1$ too, and this is indeed the case. 

The maximum of $f$ occurs when
\begin{eqnarray}
    &&\cos \phi = \frac{\Gamma_R}{\sqrt{\Gamma_R^2 + \Gamma_I^2}}\\
    &&\sin \phi = - \frac{\Gamma_I}{\sqrt{\Gamma_R^2 + \Gamma_I^2}}\\
    && \cos \theta = \sqrt{\frac{\Gamma_D^2}{\Gamma_D^2 + 4\Gamma_R^2 + 4\Gamma_I^2}} \label{eq:cos_theta}\\
    && \sin \theta = \sqrt{\frac{4\Gamma_R^2 + 4\Gamma_I^2}{\Gamma_D^2 + 4\Gamma_R^2 + 4\Gamma_I^2}}, \label{eq:sin_theta}
\end{eqnarray}
where the real constants $\Gamma_D$, $\Gamma_R$, and $\Gamma_I$ are written in terms of matrix elements of $Z_i$ as
\begin{eqnarray}
    &&\Gamma_D = \sum_{i=1}^L (Z_{\alpha\alpha} - Z_{\beta\beta})^2\\
    &&\Gamma_R + i\Gamma_I = \sum_{i=1}^L Z_{\alpha\beta}(Z_{\alpha\alpha} - Z_{\beta\beta})\label{eq:Gamma_RI}.
\end{eqnarray}
Note that the optimal rotation is always in the upper half of the Bloch sphere (see Fig.~\ref{fig:bloch_sphere}) because of the way we have associated $|a\rangle$ to $|\alpha\rangle$ and $|b\rangle$ to $|\beta\rangle$ in Eq.~\ref{eq:maximize}.

Now that we know how to ``undo'' a two-state resonance, we can compute the Floquet unitary matrix elements in the 2D basis of demixed states: 
\begin{align}
    &\begin{pmatrix}
    U_{aa} & U_{ab}\\
    U_{ba} & U_{bb}
    \end{pmatrix}
    =
     W^\dagger 
    \begin{pmatrix}
    U_{\alpha \alpha} & 0\\
    0 & U_{\beta \beta}
    \end{pmatrix}
    W, \nonumber \\
    &W= \begin{pmatrix}
    \cos(\theta/2)  & -\text{e}^{-i\phi} \sin(\theta/2)\\
    \text{e}^{i\phi} \sin(\theta/2) & \cos(\theta/2)
    \end{pmatrix} . \label{eq:2D_floquet}
\end{align}
Three relevant quantities are the size of the off-diagonal matrix element $|U_{ab}|$, the adjusted gap between the diagonal matrix elements $|U_{aa} - U_{bb}|$, and their ratio $G$~\cite{Serbyn-Abanin2015}. It follows from Eq.~\ref{eq:2D_floquet} that
\begin{eqnarray}
    &&|U_{ab}| = \frac{1}{2} |U_{\alpha\alpha} - U_{\beta\beta}| \sin \theta, \label{eq:Uab}\\
    &&|U_{aa} - U_{bb}| = |U_{\alpha\alpha} - U_{\beta\beta}| \cos \theta, \label{eq:Uaa_minus_Ubb}\\
    &&G = \frac{|U_{ab}|}{|U_{aa} - U_{bb}|} = \frac{\tan \theta}{2}, \label{eq:G_ratio}
\end{eqnarray}
where $|U_{\alpha\alpha} - U_{\beta\beta}|$ is the size of the spectral gap between the eigenstates $|\alpha\rangle$ and $|\beta\rangle$, and the cosine and sine of $\theta$ are given by Eqs.~\ref{eq:cos_theta} and \ref{eq:sin_theta}. These quantities characterize the dynamical resonance in this 2D subspace. Note that Eqs.~\ref{eq:Uab} and \ref{eq:Uaa_minus_Ubb} correctly express that as $\theta$ is increased from $0$ to its final value, the repulsion between the diagonal matrix elements $U_{aa}$ and $U_{bb}$ is decreased at the cost of generating an off-diagonal matrix element $U_{ab}$ that couples the two states. This is what we mean by ``undoing" a resonance and ``demixing" the eigenstates into their constituent localized states. The eigenstates are, by definition, not coupled by the dynamics. But they are a result of mixing states that are coupled by the dynamics, and this is our way of quantifying that idea to some extent. 

Note that we are simply looking at the same operator (the Floquet operator) in two different bases in order to identify resonances, and this is different than tuning the disorder parameter and detecting Landau-Zener-like avoided crossings in the spectrum, as was done in Ref.~\cite{Villalonga-Clark2020}. However, there is a direct connection to what was considered in Ref.~\cite{Villalonga-Clark2020} that we will now discuss: Let's imagine that we initialize a realization of our Floquet circuit at $\alpha_i = \infty$, then tune $\alpha$ down to a finite value $\alpha_f < \infty$ that we want to consider, as was done by Villalonga and Clark. Along the way, the eigenvalues of the Floquet unitary would develop some amount of (potentially very weak) all-to-all level repulsion, and they would traverse some noticeable avoided level crossings where strong many-body resonances develop. We are simply choosing to extract information about any amount of level repulsion and resonance between any given pair of states by examining the properties of the two states at $\alpha_f$, rather than examining the history of the levels from $\alpha_i$ to $\alpha_f$. In this way, we can study both very weak ``interactions" between eigenstates, as well as strong (resonant) ones, using the same rather simple approach presented in this section. 

\begin{figure}
\centering
\includegraphics[width=1.0\linewidth]{{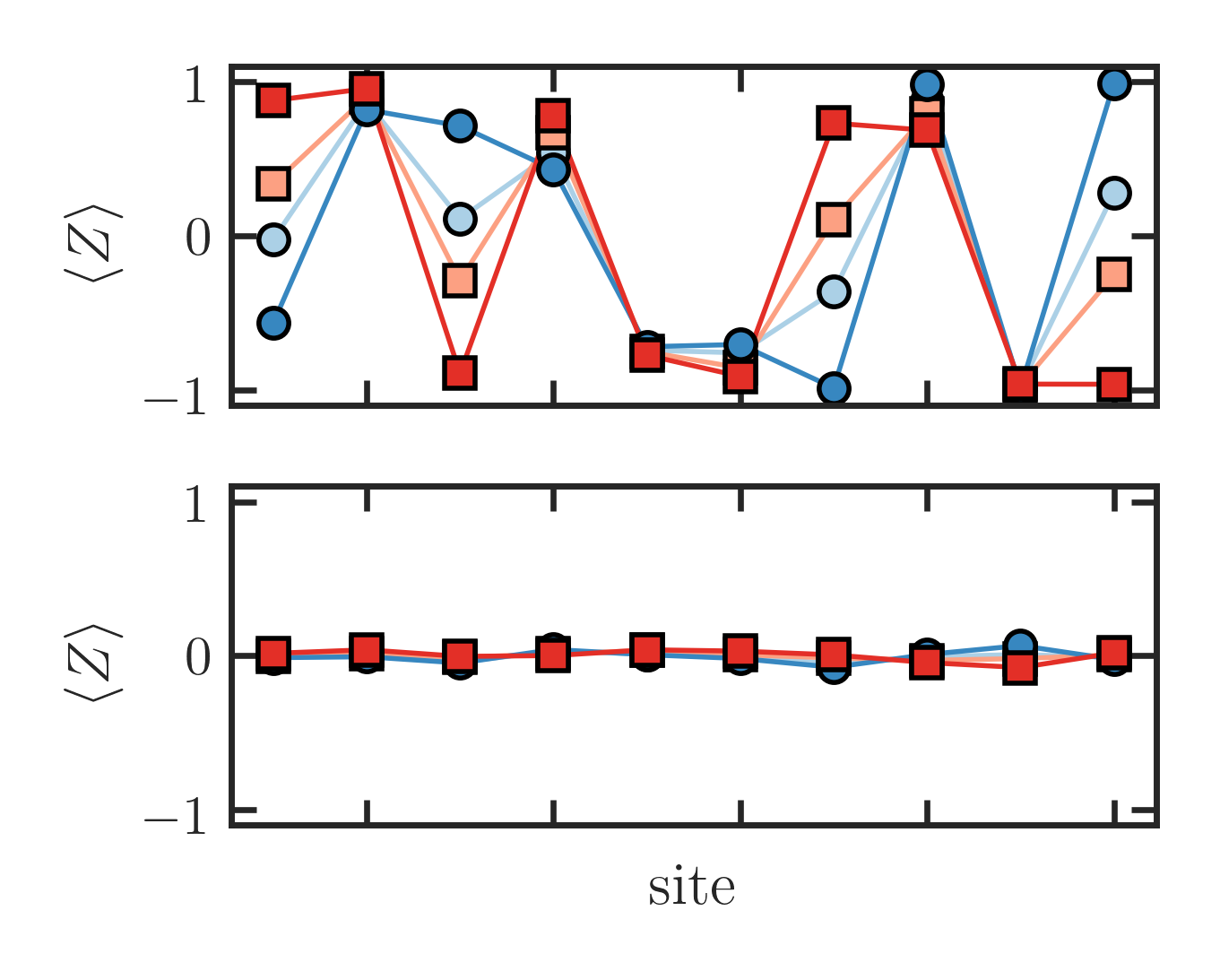}}
\caption{Demixing of two neighboring Floquet eigenstates. $Z$ magnetizations are shown for two neighboring eigenstates (light blue and red) of one realization of the Floquet model and the corresponding two demixed states (dark blue and red). The pair of neighboring eigenstates is chosen such that it is the one with the most significant rotation angle $\theta$ on the Bloch sphere (see Eq.~\ref{eq:a_state}), while also requiring that the resonance spans the system. \textbf{Top:} Floquet model at $\alpha = 12$. In a strongly localized system we find a rare two-state resonance and are able to undo the resonance by demixing the eigenstates into highly magnetized states. \textbf{Bottom:} Floquet model at $\alpha = 1$. In the thermal phase resonances involve many states, and so a pair of eigenstates in isolation are not able to be successfully demixed into highly magnetized states. \label{fig:demix}}
\end{figure}

The method we have introduced in this section is sensitive to the existence of two-state resonances, and anything weaker than that, so it works best at strong disorder where many-state resonances do not dominate the spectrum. In Fig.~\ref{fig:demix} we show the $Z$ magnetizations of two eigenstates and their corresponding demixed states in both the MBL and thermal regimes of the Floquet model. In the top panel of Fig.~\ref{fig:demix} the two eigenstates are chosen to be the two neighboring states in the spectrum of a strongly-localized system with the largest angle of rotation $\theta$ (maximum $G$), while also requiring that the resonance spans the system. We see that the demixing procedure indeed finds superpositions of the two eigenstates that are much more magnetized than the initial eigenstates. This example demonstrates the case of a well-isolated, rare, two-state resonance in the spectrum of a strongly-localized system. Meanwhile, in the bottom of Fig.~\ref{fig:demix} we do the same thing, but for a system that is well into the thermal phase. In that case the eigenstates have small initial magnetizations because they are thermal, and an attempt to find a strongly magnetized superposition of two neighboring eigenstates does not make much progress because the eigenstates are highly entangled and involved in many-state resonances.

Finally, in Fig.~\ref{fig:Uab_Hab_dist} we show distributions, over pairs of states and realizations, of the off-diagonal matrix elements $|U_{ab}|$ and $|H_{ab}|$ for various system sizes and at values of the tuning parameters $\alpha=10$ and $W=6$. Here we are restricting to pairs of eigenstates that are adjacent in the spectrum of $U$ or $H$, and for which $Z_{aa}$ and $Z_{bb}$ have opposite signs on both end sites, in order to select for pairs of states that could be a system-wide resonance (like the pair shown in the top panel of Fig.~\ref{fig:demix}). The purpose of showing these distributions is to emphasize their extreme broadness, which is growing with $L$. The distribution of off-diagonal matrix elements is much broader than the distribution of gaps, and so rare resonances are primarily caused by the tail to large off-diagonal matrix elements shown in this figure.

\section{Closed System Results\label{sec:results}}

In this section we study the two landmarks $\mathcal{L}_\text{swr}$ and $\mathcal{L}_\text{mg}$. As a reminder, these landmarks divide the MBL regime into three subregimes: Between $\mathcal{L}_r$ and $\mathcal{L}_\text{mg}$ there are rare long-range resonances, and the minimum gap does exhibit level repulsion, but the typical eigenstate is well localized and thus $\langle r \rangle \approxeq 0.39$ is near the Poisson value. Next, between $\mathcal{L}_\text{mg}$ and $\mathcal{L}_\text{swr}$ the minimum gap no longer typically exhibits level repulsion, but due to the heavy tail to large matrix elements (see Fig.~\ref{fig:Uab_Hab_dist}) the number of system-wide resonances per sample increases with increasing system size $L$.
Finally, past $\mathcal{L}_\text{swr}$ there are no system-wide resonances at all in a typical sample, and the trend with increasing system size $L$ is that samples with such resonances become even more rare.

We use extreme values, over eigenstates, of measures that indicate a many-body resonance to locate these two landmarks. 
To this end, we use our scheme for undoing two-state resonances (described in Sec.~\ref{sec:untangle}) to understand what is happening at these landmarks in terms of the matrix elements associated with many-body resonances. 
As before, we consider systems with open boundaries in order to have the longest possible distance between sites, and thus the strongest distinction between short-range and system-wide resonances in small systems. This is important because short-range resonances are certainly part of the MBL phase itself, whereas long-range (range $\sim L$) resonances are important for driving the system towards thermalizing behavior.

\subsection{System-wide resonances from long-range entanglement measures}

\begin{figure}
	\includegraphics[width=1.0\linewidth]{{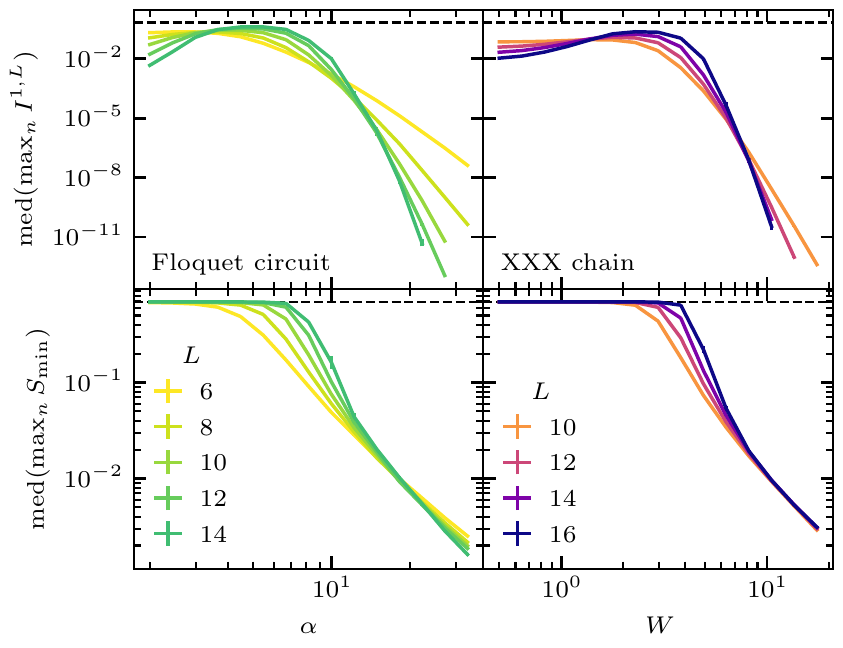}}
	\caption{End-to-end resonances: Mutual information and the maximum entanglement bottleneck. The finite-size crossings are estimates of the location of $\mathcal{L}_\text{swr}$, which we estimate to be at $\alpha > 13$ and $W > 8.5$ for $L$ larger than we can access. The dashed lines indicate the maximal value of $\ln 2$. The Floquet data is on the left and the Hamiltonian on the right.
	The Floquet data makes use of the full spectrum. The number of realizations used to compute statistics are $32 \times 10^3$ for $L \le 10$, $5 \times 10^3$ for $L=12$, and $3.5\times 10^2$ for $L=14$. 
	The Hamiltonian data is calculated from the center fifth of the spectrum in the $\sum_i Z_i = 0$ sector. The number of realizations used to compute statistics are $64 \times 10^3$ for $L\in \{10, 12\}$, $12 \times 10^3$ for $L=14$, and $5\times 10^2$ for $L=16$. 
	\textbf{Top:} The median over realizations of the maximum over eigenstates of the quantum mutual information between end sites $I^{1,L}$. The median is used in order to target typical realizations. We have dropped data points at high $\alpha$ and $W$ that are affected by finite numerical precision. \textbf{Bottom:} The median over realizations of the maximum over eigenstates of $S_\text{min}$, the minimum entanglement entropy over cuts.
	\label{fig:max_I_and_max_min_S}}
\end{figure}

Our goal is to design measures that mark the disorder strength at which we can start to expect that a typical MBL system has at least one pair of eigenstates involved in a many-body resonance that extends across the entire system ($\mathcal{L}_\text{swr}$). 
It is important to emphasize that we focus on system-wide resonances here because there is no single point at which short-range resonances turn on, but there is such a landmark for system-wide resonances, as we demonstrate below. 
In Figs.~\ref{fig:max_I_and_max_min_S} and \ref{fig:max_G_and_mean_me} we show data on four measures that are sensitive to system-wide resonances, spanning both the Floquet circuit (left) and the Hamiltonian (right) models. 

First consider Fig.~\ref{fig:max_I_and_max_min_S}, which involves two entanglement entropy-based measures of a given eigenstate, the mutual information $I^{1,L}$ between the first and last site, and $S_\text{min}$, the minimum entanglement entropy over all cuts that separate the system into left and right parts. For each eigenstate $\ket{n}$, the quantum mutual information
\begin{equation}
	I^{1,L} = S^{1} + S^{L} - S^{1,L}
	\label{eq:mutual_info}
\end{equation}
is defined by the entanglement entropies $S^{A} = -\Tr\left( \rho_A \ln \rho_A\right)$ of a subsystem $A$, which is in this case given by the first site $\{1\}$, the last site $\{L\}$, and the combination of both $\{1,L\}$. The reduced density matrix of the subsystem in eigenstate $\ket{n}$ is obtained by tracing out the complement of $A$: $\rho_A= \Tr_{\overline A} \ket{n}\bra{n}$. The minimal entanglement entropy in an eigenstate $\ket{n}$ over all cuts which separate the system into a left half $A=\{1,2,\dots, \ell\}$ and a right half $\overline A = \{\ell+1,\dots, L\}$ is then
\begin{equation}
	S_\text{min} = \min_{\ell} S^{1,2,\dots,\ell}.
\end{equation}

We then take the maximum, over eigenstates, of each of these quantities in order to get $\max_n I^{1,L}$ and $\max_n S_\text{min}$ for each realization, and finally what we plot is the median over realizations, denoted by med($\cdot$). The logic behind both of these measures is to detect, in typical realizations, rare many-body resonances that involve only a few states that differ over the entire length of the system~\cite{edge_modes}. The median is used in order to capture the typical behavior, and to postpone the influence of finite numerical precision to higher disorder. 

The quantum mutual information between end sites unambiguously picks up the small amount of system-wide entanglement in rare, system-wide, cat-like eigenstates that are the result of system-wide, few-state resonances. 
In the top row of Fig.~\ref{fig:max_I_and_max_min_S}, we see a clear crossing in the maximum mutual information at strong disorder in both models. This means that at strong disorder the end-to-end mutual information vanishes for all eigenstates as the system size (distance between ends) is increased. Before the crossing, at intermediate disorder, the median of the maximal mutual information approaches its maximal value $\ln 2$, revealing the existence of at least one system-wide resonance in most samples. So this finite-size crossing is an indicator of the location of $\mathcal{L}_\text{swr}$.
We also note that there is another crossing of this mutual information at much weaker disorder, and this must occur because in the thermal phase, the entanglement is long-range and obeys a volume law, corresponding to a small mutual information between end sites. This second crossing simply indicates that our measure is tuned to be sensitive to rare isolated resonant states in a spectrum of otherwise localized states, and not to the full thermalization of the system.

The maximal entanglement bottleneck, shown in the bottom row of Fig.~\ref{fig:max_I_and_max_min_S}, is not so unambiguous because it does not necessarily filter out short-range entanglement. If there is an eigenstate whose minimal cut corresponds to nearly one bit of entanglement entropy, this does not necessarily mean that there is system-wide entanglement in that state. However, it is true that if all states have an entanglement entropy bottleneck that is much smaller than one bit, then there are no states with significant end-to-end entanglement. Thus this quantity is useful to detect the absence of system-wide, few-state resonances at strong disorder. 

Crossings in both of the quantities in Fig.~\ref{fig:max_I_and_max_min_S}, for both Floquet and Hamiltonian systems, estimate the landmark $\mathcal{L}_\text{swr}$ beyond which not even one system-wide many-body resonance is indicated in the large-$L$ limit of a typical sample. This is not to say that resonances do not exist beyond this point, but it is to say that they do not stretch across the whole system. As in all other measures, the crossings drift towards increasing disorder strength with increasing $L$.
At larger $L$ than we can access we bound this landmark at $\alpha > 13$ for the Floquet model and $W > 8.5$ for the Hamiltonian, disorder strengths that are substantially higher than the location of $\mathcal{L}_r$ (Fig.~\ref{fig:mean_r}).

\subsection{System-wide resonances from matrix elements}

Next we move on to study resonances using the procedure outlined in Sec.~\ref{sec:untangle} for analyzing pairs of eigenstates. We restrict our analysis in this section to considering only pairs of \textit{neighboring} eigenstates for convenience (to avoid the $O(4^L)$ scaling of checking all pairs of eigenstates). We expect that system-wide resonances restricted to neighboring pairs of states ``turn on" at approximately the same disorder strength as system-wide resonances involving any pairs, and indeed our data below supports this.

\begin{figure}
\includegraphics[width=1.0\linewidth]{{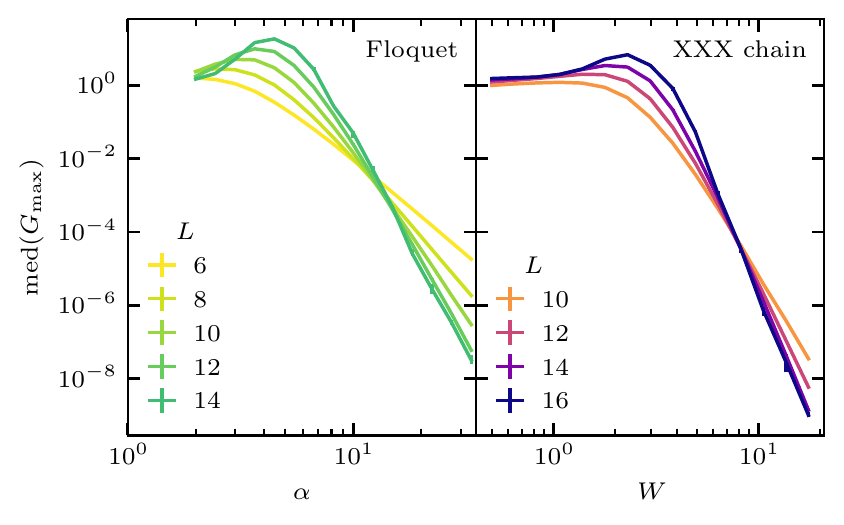}}
\caption{Typical maximum end-to-end $G$ ratio. The crossings are estimates of $\mathcal{L}_\text{swr}$. med($\cdot$) denotes the median. The median is used to target typical realizations, and the maximum is taken over pairs of neighboring eigenstates that are ``system-wide" (differ in the signs of their magnetizations on both end sites). Again the Floquet data is on the left and the Hamiltonian data is on the right. The system sizes $L$, disorder strengths $\alpha$ and $W$, and number of realizations are all the same as in Fig.~\ref{fig:max_I_and_max_min_S}. 
\label{fig:max_G_and_mean_me}}
\end{figure}

For each pair of neighboring eigenstates $(n,n+1)$ we compute the demixed states, $|a\rangle$ and $|b\rangle$, and extract the associated off-diagonal matrix element of the Floquet unitary $|U_{ab}|$ (Hamiltonian $|H_{ab}|$), the adjusted gap $|U_{aa} - U_{bb}|$ ($|H_{aa} - H_{bb}|$), and $G$, which is the ratio of the two (see Sec.~\ref{sec:untangle} for details). In order to consider only potential system-wide resonances, we further filter the pairs of states and keep only the $\sim 1/4$ of pairs for which both $\langle Z_1 \rangle$ and $\langle Z_L \rangle$ have opposite signs in the two demixed states; this is what we mean by ``system-wide" when talking about a pair of eigenstates in this section. For example, the resonance shown in the top panel of Fig.~\ref{fig:demix} is system-wide. In order to detect the onset of system-wide resonances (in neighboring pairs of eigenstates) we determine the largest $G$ from each realization, $G_\text{max} = \max_n G$, and plot the median over realizations in the top row of Fig.~\ref{fig:max_G_and_mean_me}. 
There we see that beyond a certain level of disorder, the maximum $G$ trends to lower values with $L$, and thus we do not expect to find even one such resonance in large systems at those disorders. This is in agreement with the entanglement entropy-based measures we studied above (note that those measures did not have the restriction to neighboring pairs of eigenstates).

\begin{figure}
	\includegraphics[width=1.0\linewidth]{{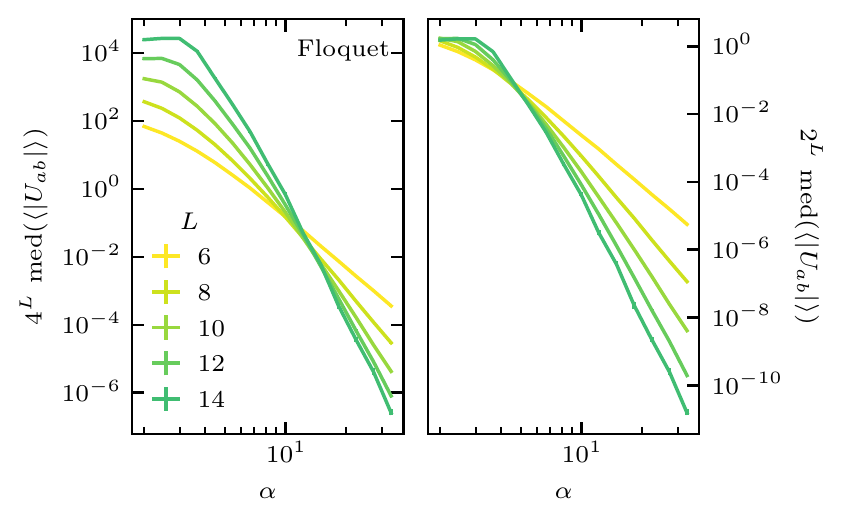}}
	\caption{Typical mean end-to-end matrix element $|U_{ab}|$, scaled by $4^L$ (left) and $2^L$ (right), for different system sizes $L$ as a function of $\alpha$ for the Floquet circuit. The crossings are estimates of $\mathcal{L}_\text{swr}$ (left) and $\mathcal{L}_r$ (right). The median is taken over realizations and the mean is taken over system-wide pairs of neighboring eigenstates. As explained in the main text, the reason we scale by $4^L$ on the left is to make a quantity that is proportional to the number of resonant system-wide pairs of neighboring eigenstates in a realization. On the right we scale the same data by $2^L$ so that the finite-size crossing occurs when the number of resonances is $\sim 2^L$, i.e., an $O(1)$ fraction of states.
	\label{fig:uab}}
\end{figure}

Now remember that for a given pair of states, $G = |U_{ab}|/|U_{aa}-U_{bb}|$ (here we focus on the Floquet model). 
We can define a criterion of two eigenstates being resonant if $G>1$, i.e., $U_{ab} > |U_{aa}-U_{bb}|$. 
In MBL systems in the regime of rare, isolated resonances in the spectrum, we can approximate $|U_{aa}-U_{bb}|$ as being distributed exponentially (Poisson statistics) with a mean value of $2\pi / 2^L$. Then if we assume that the matrix element $|U_{ab}|$ is uncorrelated to the gap $|U_{aa}-U_{bb}|$, the expected total number of resonances (restricted to system-wide neighboring pairs of states) in a realization is the sum over qualifying pairs of eigenstates of the probability that the pair satisfies the resonance condition. In the regime where the probability of resonance per pair is small enough, this number of resonances per realization is
\begin{eqnarray}
    n_\text{swr} &\approx& \sum_\text{pairs}  \int_0^{|U_{ab}|} \frac{2^L}{2\pi} \exp \left( -\frac{2^L}{2 \pi} \Delta \right) d\Delta \\ 
    &\approx& \frac{4^L}{8\pi} \langle |U_{ab}| \rangle,
\end{eqnarray}
where the mean $\langle \cdot \rangle$ is taken over neighboring system-wide pairs of states, of which there are approximately $2^L / 4$. Thus the quantity $4^L \langle |U_{ab}| \rangle $ should also herald the onset of $O(1)$ system-wide resonances involving neighboring pairs of eigenstates, and the locations of the finite-size crossings, that we show on the left of Fig.~\ref{fig:uab}, serve as additional estimates of $\mathcal{L}_\text{swr}$ (up to the assumptions of this section).
Note that a factor of $4^L$ appears several times in this work following distinct lines of reasoning, so this $4^L$ is not the same as the $4^L$ that enters in the avalanche argument discussed in Sec.~\ref{sec:avalanche_intro}. This reasoning and quantity does not translate well to the Hamiltonian case because of the nonuniform density of states, so we do not show data for this quantity for the Hamiltonian model, as we have done with the other measures.
The finite-size crossings in Figs.~\ref{fig:max_G_and_mean_me} and the left of Fig.~\ref{fig:uab} agree well with those identified above in Fig.~\ref{fig:max_I_and_max_min_S}, so we conclude that system-wide resonances involving neighboring pairs of eigenstates onset at a similar point as system-wide resonances between any pair, and we have further evidence for the landmark $\mathcal{L}_\text{swr}$.

As an extension of these ideas, on the right of Fig.~\ref{fig:uab} we set the target number of system-wide, resonant, neighboring pairs of states to $\sim 2^L$, instead of $\sim 1$, and thus plot $2^L \langle |U_{ab}| \rangle$ vs. $\alpha$. This means that the finite-size crossing occurs when the probability of a system-wide resonance in any given eigenstate is finite (does not vanish with $L$), and hence a finite fraction of eigenstates are typically involved in resonances. The location of this crossing lines up nicely with estimates of $\mathcal{L}_r$, the boundary of the MBL regime.

\subsection{Level repulsion of the minimum gap}

In this subsection we focus on a different landmark, $\mathcal{L}_\text{mg}$, where the two closest levels in the spectrum start to undergo significant level repulsion. 
We focus on the Floquet system where the smallest gap for Poisson level statistics can be obtained directly, because the density of states is known and independent of the eigenphase, but the same ideas and results hold in the Hamiltonian case too, with slight modifications. 

\begin{figure}
\includegraphics[width=1.0\linewidth]{{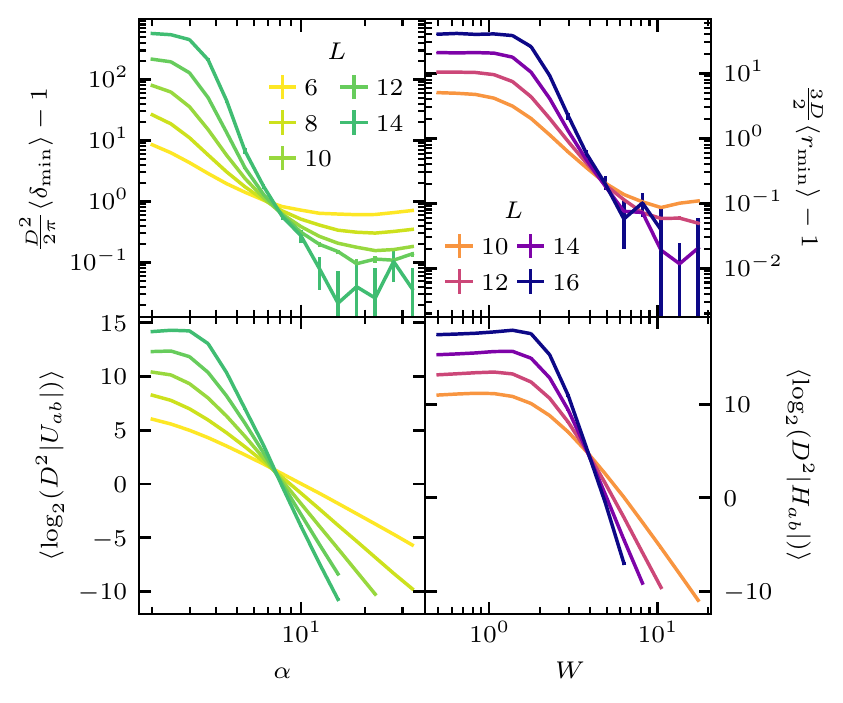}}
\caption{Level repulsion of the smallest gaps and scaling of the typical matrix element. Floquet data is on the left and Hamiltonian data is on the right, as in the above figures. The crossings are estimates of $\mathcal{L}_\text{mg}$, which we estimate to be at $\alpha > 7.9$ and $W > 5.7$ as $L$ increases. \textbf{Top:} The average minimum gap (Floquet) and minimum gap ratio (Hamiltonian). Data is scaled and shifted so that the value for a Poisson (uncorrelated) spectrum is 0 for both quantities. There is not an important difference between average and typical values for these quantities. \textbf{Bottom:} The log of the typical scaled matrix element (restricted to neighboring pairs of states, but not to system-wide pairs). $D$ is the number of states in a realization: $2^L$ for Floquet and $\frac{1}{5} \binom{L}{L/2}$ for the Hamiltonian model because we restrict to the $\sum_i Z_i = 0$ sector and take the middle fifth of eigenstates. As explained in the main text, we scale the matrix elements by $D^2$ because the minimum gap is $\sim D^{-2}$. We have dropped data points at high $\alpha$ and $W$ that are affected by finite numerical precision. \label{fig:min_gap_and_mean_log2me}}
\end{figure}

When there is no level repulsion between eigenvalues of the Floquet unitary, the probability distribution of the size of gaps $\delta$ in the spectrum is Poisson distributed
\begin{equation}
    p(\delta) = \frac{1}{\langle\delta\rangle} \exp\left(-\frac{\delta}{\langle \delta \rangle}\right),
    \label{eq:gap_pdf}
\end{equation}
with mean $\langle \delta \rangle =2\pi / 2^L$. Since there are $D=2^L$ gaps in the spectrum, the expected minimum gap is $\langle \delta_\text{min} \rangle = 2\pi / 4^L$. 
A deviation from this expectation indicates level repulsion of the smallest gap, which is caused by a many-body resonance of any range (probably not exactly end-to-end, but indeed involving extensively many degrees of freedom in order to get two levels so close together). In the top left panel of Fig.~\ref{fig:min_gap_and_mean_log2me} we show the realization-averaged minimum gap, scaled and shifted so that the baseline value for randomly placed levels is $0$. This shows that there is a landmark $\mathcal{L}_\text{mg}$ beyond which the minimum gap in the spectrum does not undergo level repulsion, and thus those eigenstates do not share a resonance. 

Assuming that for the minimum gap states, it is the gap that is atypical, and not the matrix element of the resonance in that 2D subspace, the matrix element should be typical. Thus the minimum gap should exhibit level repulsion when the typical matrix element decreases with $L$ slower than $4^{-L}$. In the bottom left of Fig.~\ref{fig:min_gap_and_mean_log2me} we show that indeed the location at which the typical value of $4^L |U_{ab}| / 2 \pi$ (over neighboring pairs of states, but not restricted to system wide pairs) has a crossing is in line with $\mathcal{L}_\text{mg}$.

On the right side of Fig.~\ref{fig:min_gap_and_mean_log2me} we show similar measures for the Hamiltonian system, however instead of the minimum gap, we use the minimum level spacing ratio in order to divide out the effect of a nonuniform density of states. The minimum level spacing ratio, $r_\text{min}$, scales with the inverse of the number of states $D$ when there is no significant level repulsion. We see that, similar to the Floquet model, there is a crossing of the minimum level spacing ratio that indicates the disorder strength at which level repulsion sets in for one of the smallest gaps in the spectrum (relative to the density of states at that energy), and this landmark is reproduced by examining when the typical matrix element between neighboring states scales like $1/D^2$ (bottom right). 

The location of $\mathcal{L}_\text{mg}$ is drifting to $\alpha > 7.9$ and $W > 5.7$. At this landmark (coming from strong disorder) the nearest levels in the spectrum begin to repel each other significantly because the typical matrix element for many-body resonances becomes large enough. By this point, atypically large matrix elements in the extremely broad distributions (see Fig.~\ref{fig:Uab_Hab_dist}) have already caused many system-wide resonances, but all of the resonances are still a vanishing fraction of the spectrum. On the way from $\mathcal{L}_\text{mg}$ to $\mathcal{L}_r$, more and more states---but still a vanishing (with $L$) fraction---exhibit level repulsion, and the crossover to the thermal regime is where finally the fraction of states involved in resonances no longer vanishes with $L$.

\begin{figure}
    \centering
    \includegraphics{{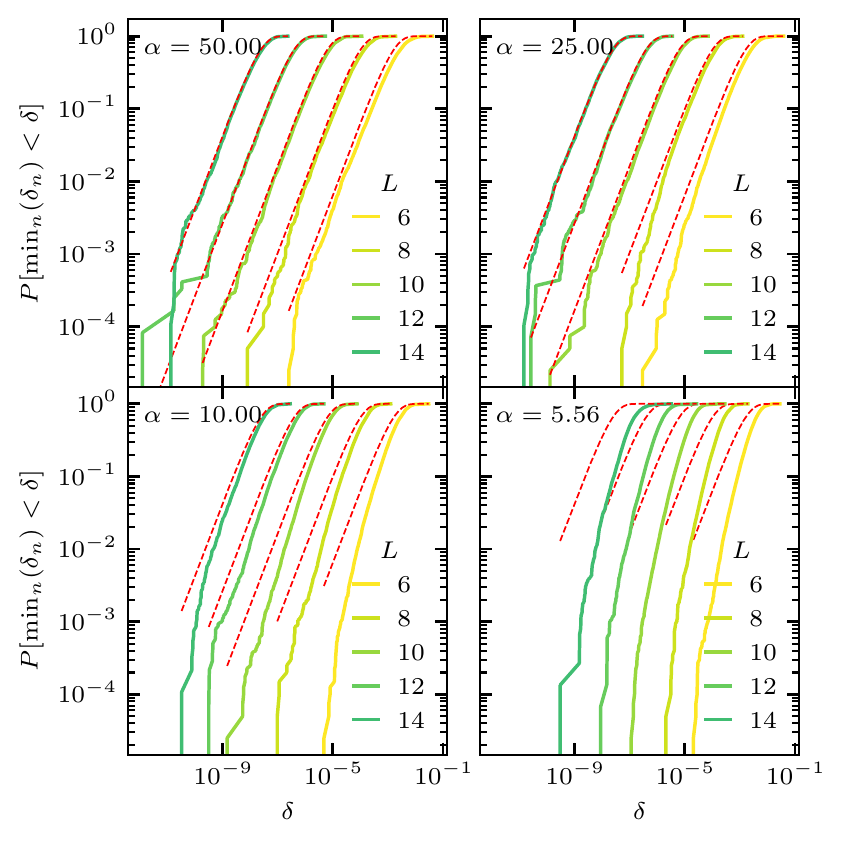}}
\caption{Cumulative probabilty density function (CDF) of the minimal gap for different system sizes $L$ and disorder strengths $\alpha$ for the Floquet unitary circuit. The dashed lines show the expected CDF $P[\text{min}_n (\delta_{n})<\delta)] = 1-\exp(4^L \delta/(2\pi))$ for the case of completely uncorrelated eigenvalues. The difference between the dashed and solid curves signals the development of level repulsion in the small gaps. \label{fig:mingap-cdf}}
\end{figure}

We can analyze the minimal gaps in more detail by considering the CDF $P[\text{min}_n(\delta_{n})<\delta)]$ shown in Fig. \ref{fig:mingap-cdf}. For the case of uncorrelated eigenvalues (Poisson statistics), we expect the distribution 
\begin{equation}
P[\text{min}_n(\delta_n)<\delta)] = 1-\exp\left(\frac{4^L \delta}{2\pi} \right),
    \label{eq:poisson-mingap-cdf}
\end{equation}
which we can compare to the numerical estimate of the CDF. The distribution in Eq. \eqref{eq:poisson-mingap-cdf} is represented by red dashed lines in Fig. \ref{fig:mingap-cdf} and is the limiting curve for all $L$ and $\alpha$, since residual level repulsion necessarily suppresses the probability to find small minimal gaps and therefore shifts the distributions to the right. It is interesting to confirm the trends already seen by the analysis of the mean of these distributions: At large disorder $\alpha\gtrsim 10$, we see that for large system sizes the uncorrelated distributions are reproduced, and the minimum gap hence does not exhibit level repulsion. On the other hand, at smaller disorder, we observe the opposite trend: for larger system sizes the observed CDF departs more strongly from the uncorrelated distribution due to level repulsion, which first occurs for small gaps.

Our data is useful to numerically check the assumption of limited level attraction in J. Imbrie's proof of the existence of MBL~\cite{imbrie_many-body_2016}. In the proof, the condition reads
\begin{equation}
P[\text{min}_n(\delta_{n})<\delta)] < \delta^\nu C^L. \label{eq:lla}
\end{equation}
It is trivially fulfilled in the case of uncorrelated levels by the CDF in Eq. \eqref{eq:poisson-mingap-cdf} with $\nu=1$ and $C=4$. Our results in Fig. \ref{fig:mingap-cdf} show compellingly that the numerical data is bounded by the Poisson CDF from above, and therefore the assumption of limited level attraction is comfortably fulfilled for all $\alpha$. We note that the target of the original proof was Hamiltonian systems, however we have considered a Floquet system here since, as discussed earlier, the spectrum is simpler to work with in some respects.

\section{Summary and discussion\label{sec:summary}}

\begin{figure}
	\includegraphics[width=\columnwidth]{{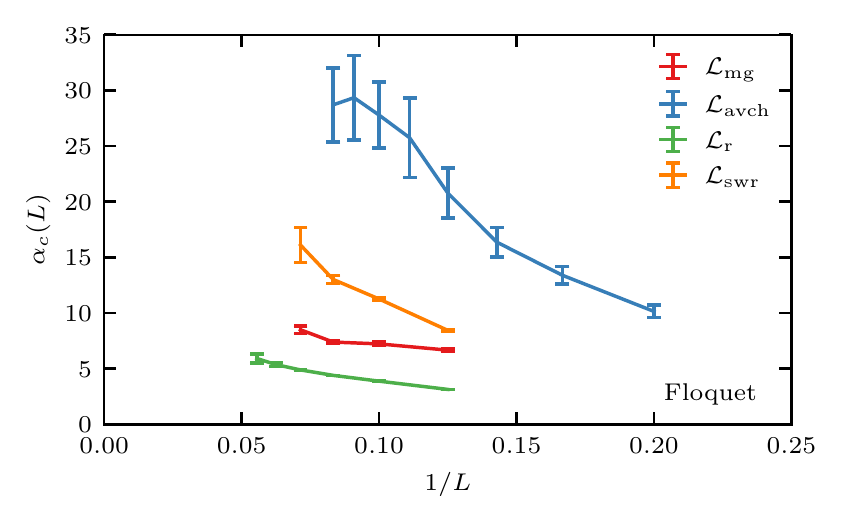}}
    \caption{Summary of estimates of landmarks in the MBL regime for the Floquet circuit. The $L$-dependent location of a landmark, $\alpha_c(L)$, is obtained via the condition $Q(\alpha_c, L) = Q(\alpha_c, L-\delta L)$, where $Q$ is a relevant quantity whose finite-size crossing indicates the landmark, and $\delta L$ is the increment by which the system size is increased for a series of data. The landmarks shown are: 1) $\mathcal{L}_\text{mg}$, where the minimal gap switches from Poissonian to non-Poissonian. $\alpha_c(L)$ is estimated from the data shown in the top left of Fig.~\ref{fig:min_gap_and_mean_log2me} with $\delta L=2$. 2) $\mathcal{L}_r$, the crossover to the thermal regime, where all states exhibit significant level repulsion. $\alpha_c(L)$ was estimated from data shown on the left of Fig.~\ref{fig:mean_r} with $\delta_L=2$. 3) $\mathcal{L}_\text{swr}$, the landmark that marks the edge of the deep-MBL subregime where no system-wide resonances are present. $\alpha_c(L)$ comes from the crossings in the top left of Fig.~\ref{fig:max_I_and_max_min_S} with $\delta L=2$. 4) $\mathcal{L}_\text{avch}$, an estimate for the position of the avalanche instability from the perturbative decay rate shown in  Fig.~\ref{fig:crossing_slowest_rate_pt_system_size}, using $\delta L=1$.
	\label{fig:landmarks} }
\end{figure}

In this work we identified and estimated several landmarks in the MBL regimes of finite-size MBL systems (see Fig~\ref{fig:landmarks}). The model systems we used are the conventional random-field Heisenberg spin chain, and a Floquet random unitary circuit that we introduced, which we argued has some simplifying advantages. In order to set reference points for these models, in Fig.~\ref{fig:mean_r} we showed the mean level spacing ratio, which is commonly used to mark the boundary of the finite-size MBL regime; that reference landmark we have called $\mathcal{L}_r$.

In addition, we defined ``open" versions of these otherwise closed spin chain models by coupling an infinite bath to the left end spin. 
The open models give us a direct handle on the avalanche instability via their superoperator description.
Based on the theory of avalanches,  the critical thermalization rate of a MBL spin chain of length $L$ coupled to an infinite bath at one end is $\sim 4^{-L}$. Identifying the slowest decaying eigenmode of the open systems as the rate of thermalization, we showed in Fig.~\ref{fig:crossing_slowest_rate} that indeed both open Floquet and Hamiltonian models have a point at which their thermalization rate crosses through the critical $~4^{-L}$ scaling. We then improved on this via a perturbative approach first suggested by Sels~\cite{sels_markovian_2021} in a follow up to the first version of this paper, which allowed us to push the calculation to larger system sizes (Fig.~\ref{fig:crossing_slowest_rate_pt_system_size}). In this way we were able to set down our first landmark $\mathcal{L}_\text{avch}$ in the MBL regimes, which is a lower-bound estimate of the boundary of the MBL \textit{phase}. Strikingly, this landmark was deep in the MBL regimes, suggesting that the boundary of the MBL phase is at much stronger randomness than indicated by studying the average properties of eigenstates/energy levels, such as $\langle r \rangle$, at accessible system sizes. It is also interesting to note that, for the Hamiltonian model, our \textit{lower} bound of the avalanche threshold ($W>18$) is beyond the \textit{upper} bound on the critical disorder strength proposed in Ref.~\cite{Geraedts-Nandkishore2017}, albeit for a slightly different model.

Before moving on to study many-body resonances in the closed spin chain models, in Sec.~\ref{sec:untangle} we detailed a new procedure for studying the effective couplings between pairs of eigenstates. The essential idea is, for any chosen pair of eigenstates, to find the two most localized superpositions of those two eigenstates, and then characterize the couplings in that 2D subspace via the matrix elements of the Floquet operator (or Hamiltonian) itself in the basis of these more localized states. Being able to study the resulting matrix elements allowed us to more thoroughly understand the landmarks that involve many-body resonances, which we identified first using other basis-independent measures. This idea of ``undoing" resonances may be useful in other settings where isolated resonances show up.

The two resonance-related landmarks we found, $\mathcal{L}_\text{swr}$ and $\mathcal{L}_\text{mg}$, split the MBL regimes into three subregimes. We detailed two ideas for entanglement entropy-based quantities that can pick up on exceedingly rare cat-like eigenstates that are involved in system-wide resonances (swr), and indeed in Fig.~\ref{fig:max_I_and_max_min_S} we showed that these worked to identify and estimate $\mathcal{L}_\text{swr}$.
We then confirmed this landmark using our procedure for undoing resonances and generating the associated matrix elements. The results are contained in Figs.~\ref{fig:max_G_and_mean_me} and \ref{fig:uab}.

The final landmark we studied was $\mathcal{L}_\text{mg}$, the point at which the minimum gap (mg) changes from being Poissonian to not.
This is the result of the typical matrix element that generates level repulsion between neighboring levels crossing through the scaling $\sim 4^{-L}$, which is how the minimum gap of a Poisson spectrum scales with $L$. This explanation was confirmed in Fig.~\ref{fig:min_gap_and_mean_log2me}.

Finally, we examined the distribution of minimum gaps as we varied $\alpha$ in our Floquet model. In Fig.~\ref{fig:mingap-cdf} we show that the distribution is bounded by the uncorrelated Poisson distribution. This is numerical confirmation, at the accessible system sizes, of the modest assumption of limited level attraction in the proof of MBL~\cite{imbrie_many-body_2016}, albeit for Floquet systems, which were not the original target of the proof.

All of our landmarks exhibit a significant drift with system size towards larger values of disorder, due to the fundamental asymmetry between thermalization and localization. Fig.~\ref{fig:landmarks} displays a summary of the landmarks and their drift with system size. It is important to note that we cannot determine whether all of our landmarks will converge as $L\to\infty$, or if any will end up in the MBL phase. 
One interesting possibility is that one of, or both of, $\mathcal{L}_\text{swr}$ and $\mathcal{L}_\text{mg}$ end up within the MBL phase, beyond $\mathcal{L}_\text{avch}$, and separate it into two or three pieces. For example, there may be a deep part of the MBL phase where typically not even a single one of the exponentially many eigenstates has any significant system-wide entanglement (compared to one bit), and then a more shallow part of the phase that does typically host these still exceedingly rare long-range resonances. Or, this may not be true, in which case it would seem that the avalanche instability would be the cause of all system-wide resonances in the limit of large samples. Our data favor the latter scenario, but we cannot rule out the former, which could occur if the character of the finite-size effects changes qualitatively at larger $L$. In the future it will be interesting to understand the connections between resonances and avalanches more thoroughly.

Our work has raised many outstanding questions for future inquiry. For example, the study of rare many-body resonances in strongly localized nonrandom (ex: quasiperiodic) MBL systems may be a new lens through which to understand the differences between random and nonrandom MBL systems~\cite{Khemani-Huse2017b}. Our method for analyzing resonances in pairs of eigenstates has also opened up the future possibility of building on this method to deal with many eigenstates at once. Furthermore, we have not yet fully understood the connection between the slowest modes of the models coupled to thermal baths and the detailed properties of the corresponding isolated systems, and this may be a challenging and rewarding direction for future work.

Soon after this paper was first posted, a related and complementary work about many-body resonances by Garratt, \textit{et al.} appeared~\cite{Garratt-Chalker2021}.

\begin{acknowledgments}

We thank Vir Bulchandani, Anushya Chandran, Phil Crowley, Sarang Gopalakrishnan, Anatoli Polkovnikov, Dries Sels, and Shivaji Sondhi for stimulating discussions. D.A.H. and A. M. were supported in part by the DARPA DRINQS program. V.K. acknowledges support from the Sloan Foundation through a Sloan Research Fellowship and by the US Department of Energy, Office of Science, Basic Energy Sciences, under Early Career Award No. DE-SC0021111. 
This work was in part supported by the Deutsche 
Forschungsgemeinschaft through SFB 1143 (project-id 247310070) and the cluster of excellence ML4Q (EXC2004, project-id 390534769).

\end{acknowledgments}

\appendix

\section{Distribution of slowest decay rates of the XXX Lindblad superoperator}
\label{app:superop-dist}

\begin{figure}
	\includegraphics[width=1.0\linewidth]{{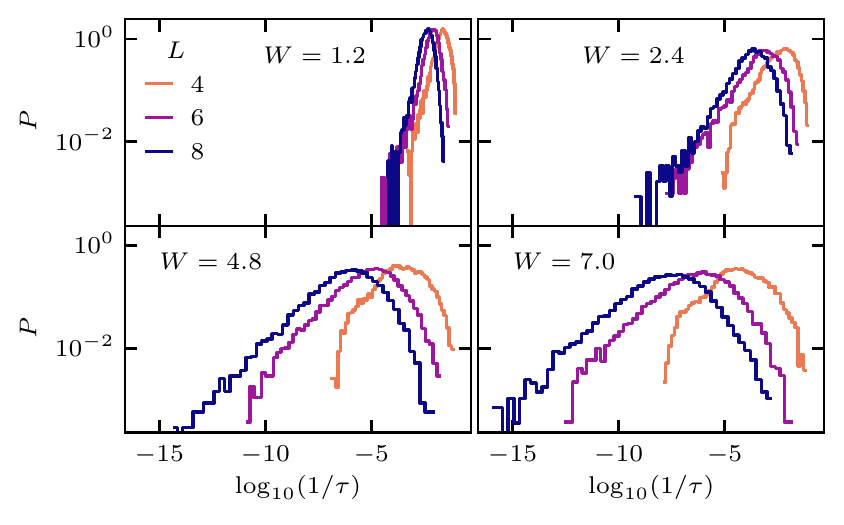}}
	\caption{Distribution, over realizations, of the slowest rate of thermalization in Lindblad dynamics for the open XXX model. This is part of the data shown in Fig.~\ref{fig:crossing_slowest_rate} \label{fig:distribution_decay_rate_xxz} }
\end{figure}

In Fig. \ref{fig:distribution_decay_rate_xxz}, we provide additional data for the distributions of the slowest rate of the Lindblad superoperator of the Hamiltonian model coupled to an infinite bath.

\section{Small matrix elements and finite numerical precision\label{app:num_precision}}
In Fig.~\ref{fig:Uab_Hab_dist} we show the distributions of the off-diagonal matrix elements obtained via our procedure for undoing resonances in the 2D subspace spanned by two chosen eigenstates. At strong disorder and large system size there is a left tail to very small numerical values, and these can be difficult to work with even for double precision (64-bit) floating point arithmetic. Small $|H_{ab}|$ are the result of small $Z_{\alpha\beta}$ in Eq.~\ref{eq:Gamma_RI}. So, the source of these issues is in the calculation of very small off-diagonal matrix elements of $Z_i$ in the basis of eigenstates.

We have found that numerical errors can distort the left tail of the distribution at numerical values of $|H_{ab}|$ near to, or smaller than, the double precision floating point resolution of $2^{-53}$. However, many statistics are not sensitive to this tail to small matrix elements and thus they are robust to some level of these errors. For example, these distributions are broad enough that the mean is dominated by the right tail to large values. The median is also robust to errors in the left tail, as long as those errors do not cause weight to be transferred from the lower to the upper half of the distribution.

\section{Numerical precision in the open setup\label{app:quad-precision}}

\begin{figure}
\includegraphics[width=1.0\linewidth]{{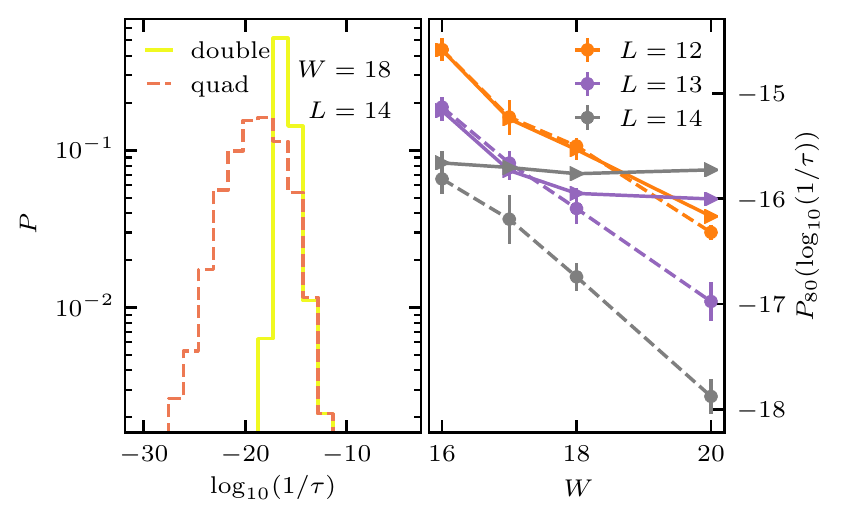}}
\caption{\textbf{Left}: Distribution of decay rate $\tau$ in the open Hamiltonian setup with $L=14$ and $W=18$ at weak coupling as described in Sec. \ref{sec:weak_coupling_hamiltonian}. Dashed (continues) lines in both panels show the decay rate computed using quadruple (double) precision. \textbf{Right}: $80^\mathrm{th}$ percentile of $-\log_{10}\tau$ as function of disorder $W$ at fixed system sizes. In both plots 1000 disorder realizations are used for every set of parameters. At each disorder realization the decay rate is computed twice, once using double precision and the other using quadruple precision. Error bars are 68\% bootstrap confidence interval  \label{fig:quad_precision}}
\end{figure}

Numerical precision issues are also present in the open setups (both Floquet and Hamiltonian) when the decay rate distribution has significant support on values smaller than $10^{-16}$. Unlike the case of matrix elements coming from ``undoing" many body resonances, distributions of decay rate can lay in the region were numerical precision issues can not be ignored. This can be seen clearly in Fig.~\ref{fig:quad_precision}, where the decay rate computed using double and quadruple precision are put side by side. After looking at the distribution of decay rates, it is evident that the median of $\log_{10} (1/\tau)$ is totally changed by the insufficiency of double precision. However higher percentile of the distribution are less affected by this issue because the right tail will be at values larger than $10^{-16}$. That is why we have chosen (inspired by \cite{sels_markovian_2021}) to work with the $80^\mathrm{th}$ percentile. Moreover, different percentiles have similar system size scaling because the variance of distribution is independent of system size (see Fig. \ref{fig:distribution_decay_rate_xxz} and \ref{fig:distribution_decay_rate_floquet}). Therefore the $80^\mathrm{th}$ percentile still carries enough information for talking about ``typical" behavior. Still at large disorder and available system sizes the decay rates are too small for getting a trustworthy result out of double precision computations. For instance, in Fig. \ref{fig:quad_precision} it can be seen that for $L=14$ the $80^\mathrm{th}$ percentile of the decay rate is smaller than $10^{-16}$ in the range $W>16$. To overcome this issue we have performed quadruple precision diagonalization of $H$ and $U$ as well as their respective super-operators (see Sec. \ref{sec:weak_coupling_hamiltonian} and \ref{sec:weak_coupling_floquet}) anytime the needed percentile of the distribution is smaller than $10^{-15}$.
 
\bibliography{main}

\begin{thebibliography}{84}
\bibitem [{edg()}]{edge_modes}%
  \BibitemOpen
  \href@noop {} {}\bibinfo {note} {Systems with topological order may have
  eigenstates with end-to-end entanglement between edge modes that does not
  involve a bulk resonance~\cite{Fendley_2016}. But the models we study in this
  work do not have any such special edge modes, so the resonances that we study
  here are indeed system-wide resonances that also involve the bulk degrees of
  freedom.}\BibitemShut {Stop}%
\end{thebibliography}

\end{document}